\documentclass[epj,nopacs]{svjour}
\usepackage{graphics}
\usepackage{amsmath}
\usepackage{amssymb}
\usepackage{textcomp}
\usepackage[colorlinks=true,linktocpage=true,linkcolor=blue,citecolor=blue]{hyperref}
\usepackage[utf8x]{inputenc}
\usepackage[english]{babel}
\usepackage[babel]{csquotes}
\usepackage{indentfirst}
\usepackage{array}
\usepackage{epsfig}
\usepackage{graphicx}
\usepackage{latexsym}
\usepackage{array}
\usepackage{fancyhdr}
\usepackage{pstricks}
\usepackage{bm}
\usepackage{slashed}
\usepackage{color}
\usepackage{booktabs}
\usepackage{braket}
\usepackage{cite}
\usepackage{tabularx}
\usepackage{multirow}
\usepackage{booktabs}
\newcommand{\eq}{Eq.~}
\newcommand{\eqs}{Eqs.~}

\newcommand{\reff}{Ref.~}

\newcommand{\secc}{Sec.~}
\newcommand{\app}{App.}
\newcommand{\rr}{\hat{\rho}}
\def\wT{{\hat T}}

\newcommand{\tr}{{\rm tr}}

\newcommand{\Tr}{\text{Tr}}
\newcommand{\n}{\nonumber}

\newcommand{\F}{\mathcal{F}}
\newcommand{\Pc}{\mathcal{P}}
\newcommand{\V}{\mathcal{V}}
\newcommand{\A}{\mathcal{A}}
\newcommand{\mC}{\mathfrak{C}}
\newcommand{\C}{\mathcal{C}}
\newcommand{\Sc}{\mathcal{S}}
\newcommand{\ms}{\mathfrak{s}}
\newcommand{\f}{\mathfrak{f}}

\newcommand{\beq}{\begin{equation}}
\newcommand{\eeq}{\end{equation}}
 \newcommand{\im}{\text{Im}\, }
 \newcommand{\re}{\text{Re}\, }

 \newcommand{\olra}{\overleftrightarrow}
 \newcommand{\ps}{\psi}
 \newcommand{\psbar}{\bar{\psi}}
 \newcommand{\T}{\hat{T}}
 \newcommand{\Sp}{\hat{S}}
  \newcommand{\J}{\hat{J}}
    \newcommand{\nn}{N}
    \newcommand{\pp}{\boldsymbol{\mathfrak{p}}}
    \newcommand{\ppm}{\mathfrak{p}}
  \newcommand{\nnn}{N}
  \newcommand{\trans}{\text{pgt}}
\numberwithin{equation}{section}
\begin{document}
\title{Spin tensor and pseudo-gauges: from nuclear collisions to gravitational physics}
%\subtitle{}
\author{Enrico Speranza
%\thanks{\emph{Present address:} Insert the address here if needed}
\and Nora Weickgenannt% etc
% \thanks is optional - remove next line if not needed
%\thanks{\emph{Present address:} Insert the address here if needed}
%
}                     % Do not remove
%
%\offprints{}          % Insert a name or remove this line
%
\institute{Institute for Theoretical Physics, Goethe University,
Max-von-Laue-Str. 1, D-60438 Frankfurt am Main, Germany}
\date{}
% The correct dates will be entered by Springer
%
\abstract{
The relativistic treatment of spin is a fundamental subject which has an old history. In various physical contexts it is necessary to separate the relativistic total angular momentum into an orbital and spin contribution. However, such decomposition is affected by ambiguities since one can always redefine the orbital and spin part through the so-called pseudo-gauge transformations. We analyze this problem in detail by discussing the most common choices of energy-momentum and spin tensors with an emphasis on their physical implications, {and study the spin vector which is a pseudo-gauge invariant operator}.  We review the angular momentum decomposition as a crucial ingredient for the formulation of relativistic spin hydrodynamics and quantum kinetic theory with a focus on relativistic nuclear collisions, where spin physics has recently attracted significant attention. Furthermore, we point out the connection between pseudo-gauge transformations and the different definitions of the relativistic center of inertia.
Finally, we consider the Einstein-Cartan theory, an extension of conventional general relativity, which allows for a natural definition of the spin tensor. 
%
%\PACS{
%      {PACS-key}{discribing text of that key}   \and
%      {PACS-key}{discribing text of that key}
%     } % end of PACS codes
} %end of abstract
\setcounter{tocdepth}{2}
\maketitle
\tableofcontents

\section{Introduction}
\label{intro}

The decomposition of the relativistic total angular momentum into an orbital and spin part is a long-standing problem both in quantum field theory and gravitational physics~\cite{Hehl:1976vr}. The definitions of the energy-momentum and spin tensors used to construct the total angular momentum density suffer from ambiguities. In fact, one can always redefine them through the so-called pseudo-gauge transformations such that the total charges (i.e., the total energy, momentum and angular momentum) do not change. As long as one is only interested in the total charges, this ambiguity is clearly of no importance. {However, in many physical contexts, it is important to separate the orbital and spin angular momentum of the system. The question we would like to address can be stated as follows: is there a particular choice for the angular momentum decomposition for a specific system which gives a ``physical" local distribution of energy, momentum and spin? Here, by ``physical" we mean that such decomposition can have some consequences on experimental observables. Although in conventional Einstein's general relativity the answer to this question is that the energy-momentum density is measurable through geometry, this issue is currently debated in quantum field theory where spin degrees of freedom are considered. We stress that, in this paper, we do not aim at finding a physical decomposition which is valid for all possible systems. Rather, 
we address this problem by studying different contexts, indicating for each of them the implications of various pseudo-gauge choices and which decomposition appears to be a physical one. 
}

The paper is structured as follows. In \secc\ref{secvarious} we introduce the basic concepts: the spin tensor and pseudo-gauge transformations, and discuss some of the most common choices in the literature. In \secc\ref{pl_sec}, we generalize the concept of nonrelativistic spin vector to the relativistic case, discussing the Frenkel theory and the Pauli-Lubanski pseudovector. In \secc\ref{wigsec} we consider the Wigner-operator formalism.
Furthermore, in \secc\ref{nonsec} we  study the effect of different pseudo-gauge choices in thermodynamics using the method of Zubarev to construct the statistical operator. {We show that the local equilibrium state is in general pseudo-gauge dependent, unlike the global equilibrium one. This leads to the consequence that expectation values of observables are in general sensitive to different pseudo-gauges for many-body systems away from global equilibrium.} 
In relativistic heavy-ion collisions, spin-pola{\-}rization phenomena related to medium rotation have recently attracted significant interest. This is due to experimental observations showing that certain hadrons emitted in noncentral collisions are indeed produced with a finite spin polarization \cite{STAR:2017ckg}. The Wigner operator turns out to be particularly suitable to study spin effects in heavy-ion collisions and a brief review on some recent applications in the field is given in \secc\ref{hicsec}.  In \secc\ref{centsec} we make a connection between  pseudo-gauges and the relativistic center of mass which, unlike in the nonrelativistic case, suffers from ambiguities in the definition. {In particular, we show that the redefinition of the center of inertia can be related to a different decomposition of orbital and spin angular momentum of a relativistic system which, in turn, is associated to a specific spin vector \cite{Pryce:1948pf}.} In the end, in \secc\ref{grsec} the physical meaning of the energy-momentum tensor in general relativity is discussed. We outline an extension of conventional general relativity, called Einstein-Cartan theory, where a natural definition of spin tensor arises by allowing the spacetime to a have a nonvanishing torsion. 
Brief conclusions are given in \secc\ref{concsec}.

We use the following notation and conventions: $a\cdot b=a^\mu b_\mu$,
$a_{[\mu}b_{\nu]}\equiv a_\mu b_\nu-a_\nu b_\mu$, $g_{\mu \nu} = \mathrm{diag}(+,-,-,-)$ for the Minkowski metric, and
$\epsilon^{0123} = - \epsilon_{0123} = 1$.

 \section{Spin tensor and pseudo-gauge transformations}
 \label{secvarious}
 
The spin tensor is one of the fundamental quantities we will consider in this paper. In this section we give a definition starting from Noether's theorem and discuss the pseudo-gauge transformations which allow a redefinition of energy-momentum and spin tensors. We review some of the most commonly used pseudo-gauge choices in the literature and explore their physical implications. In this paper we will focus on the Dirac theory.
 
  \subsection{Canonical currents}
  \label{seccan}
  
Let us consider the Lagrangian density for the free Dirac field $\psi(x)$ with mass $m$
  \begin{equation}
  \label{ld}
   \mathcal{L}_D(x)=\frac {i\hbar} 2 \,  \psbar (x)\gamma^\mu\olra{\partial}_\mu\ps (x) - m\psbar (x)\ps (x),
  \end{equation}
  where $\olra{\partial}^\mu\equiv \overrightarrow{\partial}^\mu-\overleftarrow{\partial}^\mu$ and $\gamma^\mu$ are the Dirac matrices. 
  The corresponding action is given by
 \begin{equation}
 \label{action}
 A=\int d^4 x \, \mathcal{L}_D(x) .
 \end{equation}
 The equations of motion associated to the Lagrangian \eqref{ld} are the Dirac equation for the field and its adjoint
  \begin{subequations}
  \label{diraceq}
  \begin{align}
  \label{diraceq1}
  (i\hbar\gamma^\mu \partial_\mu - m)\ps (x)&=0, \\
  \psbar (x)(i\hbar\gamma^\mu \overleftarrow{\partial}_\mu + m)&=0, \label{diraceq2}
 \end{align}
  \end{subequations}
  respectively.
Consider the infinitesimal spacetime translations 
\begin{equation}
\label{stt}
x^\mu \to x^{\prime\mu}=x^\mu + \xi^\mu,
\end{equation}
with $\xi^\mu$ a constant parameter.
The canonical energy-mo{\-}mentum tensor $\T_C^{\mu\nu}(x)$ is defined as the conserved current obtained using Noether's theorem by requiring the invariance of the action 
 under the transformations in Eq. \eqref{stt} \cite{DeGroot:1980dk,weinberg1995quantum}:
  \footnote{For the sake of ease of notation we will omit the $x$ dependence when there is no risk of confusion.}
 \begin{equation}
 \label{c1}
 \partial_\mu \T^{\mu\nu}_C=0,
 \end{equation}
 where
\begin{align}
\T^{\mu\nu}_C&=\frac{\partial \mathcal{L}_D}{\partial (\partial_\mu \ps)}\partial^\nu \ps + \partial^\nu \psbar \frac{\partial \mathcal{L}_D}{\partial (\partial_\mu \psbar)}-g^{\mu\nu}\mathcal{L}_D  \n\\
&= \frac{i\hbar}{2}\psbar\gamma^\mu \olra{\partial}^\nu\ps - g^{\mu\nu}\mathcal{L}_D .
\label{tcan}
\end{align}
Note that $\T^{\mu\nu}_C$ is not symmetric.
Equation \eqref{c1} implies that, for any three-dimensional space-like hypersurface $\Sigma_\lambda$, the total four-momentum operator is given by
\begin{equation}
\label{totp}
\hat{P}^\mu=\int_\Sigma d \Sigma_\lambda\, \T_C^{\lambda\mu} = i\hbar\int d^3 x \, \ps^\dagger \partial^\mu \ps
\end{equation}
where we assumed that boundary terms vanish and used the Dirac equation in Eq. \eqref{diraceq}. {In the second equality we chose the hyperplane at constant $x^0$. Since $\T_C^{\mu\nu}$ is conserved, the total charge is independent of the choice of the hypersurface integration and it transforms properly as a four-vector under Lorentz transformations, for the proof see \app \ \ref{applor}.}
 It is worth mentioning that the four components of $\hat{P}^\mu$ coincide with the four generators of the spacetime translations~\cite{weinberg1995quantum}.
The operatorial structure of $\hat{P}^\mu$ derives from the creation and annihilation operators inside the quantized fields $\ps$ and $\ps^\dagger$. The action of $\hat{P}^\mu$ on a one-particle state $\ket{\ppm, s}$, $\ppm$ being the particle four-momentum and $s$ the spin projection, is such that \footnote{In order to avoid divergences due to the vacuum expectation values, we implicitly assume the normal ordering of the operators.}
\begin{equation}
\label{poi}
\hat{P}^\mu \ket{\ppm, s} = \ppm^\mu  \ket{\ppm, s}
\end{equation}
(we use the symbol $\ppm^\mu$ for the particle momentum to distinguish it from the momentum variable $p^\mu$ of the Wigner operator discussed in \secc\ref{wigsec}).

Under the action of infinitesimal Lorentz four-rotations
\begin{equation}
\label{ptlor}
x^\mu \to x^{\prime\mu}=x^\mu + \zeta^{\mu\nu}x_\nu,
\end{equation}
with $\zeta^{\mu\nu}=-\zeta^{\nu\mu}$ constant, the total variation of the spinor reads
\begin{equation}
\delta_T\ps = \ps^\prime(x^\prime)-\ps(x)=\frac12 \zeta_{\mu\nu}f^{\mu\nu}\ps (x) ,
\end{equation}
with $f^{\mu\nu}=-\frac i2 \sigma^{\mu\nu}$ and
\begin{equation}
\label{sig}
\sigma^{\mu\nu}=\frac i2[\gamma^\mu, \gamma^\nu].
\end{equation}
Using Noether's theorem, the invariance of the action under the transformations \eqref{ptlor} yields the conservation of the canonical total angular momentum tensor $\J^{\lambda, \mu\nu}_C$ \cite{DeGroot:1980dk,weinberg1995quantum},
  \begin{equation}
  \label{c2}
  \partial_\lambda \J^{\lambda, \mu\nu}_C =0.
  \end{equation}
  where
  \begin{equation}
  \label{js}
  \J^{\lambda, \mu\nu}_C = x^\mu \T_C^{\lambda\nu}-x^\nu \T_C^{\lambda\mu} + \hat{S}_C^{\lambda,\mu\nu}.
  \end{equation}
 The tensor $\Sp_C^{\lambda,\mu\nu}$ is called canonical spin tensor and it is defined as 
 \begin{align}
 \label{scan}
 \Sp_C^{\lambda,\mu\nu}&= \frac{\partial \mathcal{L}_D}{\partial (\partial_\lambda \ps)}f^{\mu\nu} \ps -  \psbar f^{\mu\nu} \frac{\partial \mathcal{L}_D}{\partial (\partial_\lambda \psbar)}  \n \\
 &= \frac\hbar 4\psbar\{\gamma^\lambda,\sigma^{\mu\nu}\}\ps \n\\
 &=-\frac \hbar 2 \epsilon^{\lambda\mu\nu\alpha}\psbar\gamma_\alpha \gamma_5 \ps .
\end{align}  
 Inserting \eq\eqref{js} into \eqref{c2}, we obtain a (non)conser{\-}vation law for the spin tensor
\begin{equation}
\label{nccan}
\partial_\lambda \Sp_C^{\lambda,\mu\nu}=\T_C^{[\nu\mu]}.
\end{equation}
Note that, since $\T_C^{\mu\nu}$ is not symmetric, the spin tensor is not conserved. The conserved charge associated to Eq. \eqref{c2} is the total angular momentum
\begin{align}
\label{totj}
\J^{\mu\nu}&=\int_\Sigma d \Sigma_\lambda \,\J_C^{\lambda,\mu\nu}
 \n \\
  &=\int d^3x\, \ps^\dagger \left[ i\hbar(x^\mu \partial^{\nu}-x^\nu \partial^\nu) + \frac \hbar 2 \sigma^{\mu\nu}\right]\ps ,
\end{align} 
{where the hypersurface integration at constant $x^0$ was used. Since $\hat{J}_C^{\lambda,\mu\nu}$ is conserved, the quantity $\J^{\mu\nu}$ is independent of the hypersurface and it transforms 
as a rank-two antisymmetric tensor (see \app \ \ref{applor}). Moreover, the six independent components of $\J^{\mu\nu}$ are the generators of the Lorentz transformations~\cite{weinberg1995quantum}. }

By looking at \eqs\eqref{js} and \eqref{totj}, it would 
be suggestive to identify the terms involving $\T_C^{\mu\nu}$ as the orbital angular momentum density and $\Sp_C^{\lambda,\mu\nu}$ as the spin density. In fact, for a scalar field the spin tensor vanishes and we are left only with an orbital angular momentum-like contribution, so that this intuitive interpretation of $\Sp_C^{\lambda,\mu\nu}$ as a spin density would seem to work. However, there are two reasons why this identification cannot be done straightforwardly. 
The first reason is that the quantity that we call global spin defined as
 \begin{equation}
 \label{totcanspin}
  \Sp^{\mu\nu}_C\equiv \int_\Sigma d \Sigma_\lambda\, \Sp_C^{\lambda,\mu\nu}
 \end{equation}
{is not a Lorentz tensor since, as can be seen from Eq. \eqref{nccan}, $\Sp_C^{\lambda,\mu\nu}$ is not conserved (see \app \ \ref{applor}). This also means that the canonical global spin is not independent of the choice of the hypersurface.} 
Hence, the usage of the canonical spin tensor does not lead to a covariant description of spin for free fields which, instead, is something one would like to require. We will discuss a solution to this problem in Secs. \ref{HWsec}, \ref{KGsec} and \ref{GLWsec}. 
The second reason is that the decomposition of orbital and spin angular momentum in \eq\eqref{js} is ambiguous. It is indeed always possible to define a new pair of tensors $\T_\trans^{\mu\nu}$ and $\Sp_\trans^{\lambda,\mu\nu}$ connected to the canonical currents through the so-called pseudo-gauge transformations~\cite{belinfante1939spin,belinfante1940current,rosenfeld1940energy,Hehl:1976vr}
   \begin{subequations}
   \label{pgt}
   \begin{align}
   \label{pgt1t}
\T_\trans^{\mu\nu}&= \T_C^{\mu\nu}+\frac{1}{2}\partial_{\lambda}(\hat{\Phi}^{\lambda,\mu\nu}+\hat{\Phi}^{\nu,\mu\lambda}+\hat{\Phi}^{\mu,\nu\lambda}), \\
  \Sp_\trans^{\lambda,\mu\nu}&= \Sp_C^{\lambda,\mu\nu}-\hat{\Phi}^{\lambda,\mu\nu}+ \partial_\rho \hat{Z}^{\mu\nu,\lambda\rho}.
  \label{pgt1s}
  \end{align}
 \end{subequations}
{The quantities $\hat{\Phi}^{\lambda,\mu\nu}$ and $\hat{Z}^{\mu\nu,\lambda\rho}$ are arbitrary differentiable operators called superpotentials satisfying $\hat{\Phi}^{\lambda,\mu\nu}=-\hat{\Phi}^{\lambda,\nu\mu}$ and $\hat{Z}^{\mu\nu,\lambda\rho}=-\hat{Z}^{\nu\mu,\lambda\rho}=-\hat{Z}^{\mu\nu,\rho\lambda}$. It is easy to check that the new tensors $\T_\trans^{\mu\nu}$ and $\J_\trans^{\lambda,\mu\nu}=x^\mu\T_\trans^{\lambda\nu} - x^\nu\T_\trans^{\lambda\mu} +  \Sp_\trans^{\lambda,\mu\nu}$ defined through \eqs\eqref{pgt}, lead to the same total charges as in \eqs\eqref{totp} and \eqref{totj}, but in general different global spin, once integrated over some hypersurface and provided that boundary terms vanish. In the case of pseudo-gauge transformations with the same $\hat{\Phi}^{\lambda,\mu\nu}$ and different $\hat{Z}^{\mu\nu,\rho\lambda}$, the new global spins coincide. Furthermore, the new currents
are also conserved, 
\begin{equation}
\partial_\mu \T_\trans^{\mu\nu }=0, \qquad \partial_\lambda \J_\trans^{\lambda, \mu\nu} =0,
\end{equation}
 like for the canonical currents in \eqs\eqref{c1} and \eqref{c2}.
We note that, if we consider a pseudo-gauge transformation leading to a symmetric energy-momentum tensor, i.e., $\T_\trans^{[\mu\nu]}=0$, the superpotential $\hat{\Phi}^{\lambda,\mu\nu}$ is not completely arbitrary, but constrained by the relation
\begin{equation}
\partial_\lambda \hat{\Phi}^{\lambda,\mu\nu} = \T_C^{[\nu\mu]} ,
\end{equation}
as can be seen from \eq\eqref{pgt1t}. 
}

Following \reff\cite{Hehl:1976vr}, the viewpoint adopted in this paper is to assign the physical meaning of energy, momentum and spin densities to the energy-momentum and spin tensors. 
Moreover, we assume that such densities can in principle be measurable quantities. 
The act of performing the pseudo-gauge transformations can be understood as ``relocalization" of energy, momentum and spin. {The goal is then to find out which choice of the superpotentials gives a ``physical" pair of tensors to be used in a specific context.} In the next subsections we discuss several choices of pseudo-gauge transformations.

\subsection{Belinfante-Rosenfeld currents}
\label{Bel:sec}

As mentioned above, the canonical energy-momentum tensor~\eqref{tcan} is not symmetric. This inevitably leads to a conceptual problem in conventional general relativity where the energy-momentum tensor is assumed to be symmetric because defined as the variation of the action with respect to the metric, see a related discussion in \secc\ref{grsec}. This issue was overcome by Belinfante and Rosenfeld~\cite{belinfante1939spin,belinfante1940current,rosenfeld1940energy}.
It is possible to perform a pseudo-gauge transformation \eqref{pgt} with the superpotentials given by
\begin{equation}
\label{spbel}
\hat{\Phi}^{\lambda,\mu\nu}=\Sp^{\lambda,\mu\nu}_C, \qquad \hat{Z}^{\mu\nu,\lambda\rho}=0,
\end{equation}
such that the new energy-momentum tensor is symmetric (corresponding to the symmetric part of the canonical energy-momentum tensor $\Sp^{\lambda,\mu\nu}_C$) and the new spin tensor vanishes, i.e.,
\begin{align}
\label{tbel}
\T^{\mu\nu}_B &= \frac{i\hbar}{4}\psbar (\gamma^\mu \olra{\partial}^\nu + \gamma^\nu \olra{\partial}^\mu) \ps - g^{\mu\nu}\mathcal{L}_D, \\
\Sp^{\lambda,\mu\nu}_B &= 0 .
\label{sbel}
\end{align}
Consequently, the angular momentum tensor can be cast in a purely orbital-like form.

We report in passing that in the original works by Belinfante~\cite{belinfante1939spin,belinfante1940current} the approach is slightly different.
One can actually define a Belinfante spin tensor $\hat{\bar{S}}^{\lambda,\mu\nu}_B$ by decomposing the Belinfante angular momentum in the following way
\begin{align}
\J^{\lambda,\mu\nu}_B &= x^\mu\T^{\lambda\nu}_B - x^\nu\T^{\lambda\mu}_B \n\\
&= x^\mu\T^{\lambda\nu}_C - x^\nu\T^{\lambda\mu}_C+ \hat{\bar{S}}^{\lambda,\mu\nu}_B,
\label{bj}
\end{align}
with
\begin{equation}
\hat{\bar{S}}^{\lambda,\mu\nu}_B=\frac 1 2 \Big( x^\mu \partial_\rho \Sp_C^{\rho,\lambda \nu} -  x^\nu \partial_\rho \Sp_C^{\rho,\lambda \mu}\Big).
\end{equation}
However, in this paper, by Belinfante spin tensor we always mean the vanishing one~\eqref{sbel}.

\subsection{Hilgevoord-Wouthuysen currents}
\label{HWsec}

In this section we show a different decomposition between spin and orbital angular momentum first propsed by Hilgevoord and Wouthuysen (HW) such that the global spin transforms properly as a tensor under Lorentz transformations, unlike the canonical one~\cite{hilgevoord1963spin,hilgevoord1965covariant,fradkin1961tensor}. The idea is based on the fact that the Dirac spinor is also a solution of the Klein-Gordon equation. The strategy is to start from the Klein-Gordon Lagrangian, derive the currents through Noether's theorem and use the Dirac equation as a subsidiary condition. It follows that these currents will also be conserved for the Dirac theory.

The Klein-Gordon Lagrangian for the spinor reads
 \begin{equation}
 \label{lkg}
  \mathcal{L}_{KG}=\frac{1}{2m}(\hbar^2\partial_\mu\psbar \partial^\mu\ps-m^2\bar{\psi}\psi) 
 \end{equation}
 and the corresponding equations of motion are
 \begin{equation}
  \label{kgeqs}
( \hbar^2 \partial_\mu \partial^\mu + m^2)\ps=0, \quad
( \hbar^2 \partial_\mu \partial^\mu + m^2)\psbar=0.
 \end{equation}
By requiring the invariance of the action under spacetime translations and Lorentz four-rotations, we obtain the Klein-Gordon canonical energy-momentum and spin tensors using Noether's theorem. These are respectively given by the current defined in the first line of \eq\eqref{tcan}, with $\mathcal{L}_{KG}$ instead of $\mathcal{L}_{D}$, and by \eq\eqref{scan}:
\begin{subequations}
\begin{align}
\label{thw}
  \T^{\mu\nu}_{HW}&=\frac{\hbar^2}{2m}\left( \partial^\mu\psbar\partial^\nu\ps + \partial^\nu\psbar \partial^\mu\ps\right)
  -g^{\mu\nu}\mathcal{L}_{KG}, \\
 \Sp_{HW}^{\lambda,\mu\nu} &= \frac{i\hbar^2}{4m}\psbar\sigma^{\mu\nu}\olra{\partial}^\lambda \ps .
 \label{shw}
\end{align}
\end{subequations}
Notice that, in contrast to the canonical energy-momentum tensor derived from the Dirac Lagrangian, $ \T^{\mu\nu}_{HW}$ is symmetric, then 
\begin{equation}
\partial_\lambda \Sp_{HW}^{\lambda,\mu\nu} =0,
\end{equation}
which also follows after using the Klein-Gordon equations \eqref{kgeqs}.
We now require that $\psi$ is also a solution of the Dirac equation. Multiplying~\eqref{diraceq1} and   \eqref{diraceq2} by $\gamma^\lambda$ on the left and on the right, respectively, and using the identity $\gamma^\lambda \gamma^\mu=g^{\lambda\mu}-i\sigma^{\lambda\mu}$, the Dirac equations can be written as
\begin{subequations}
\label{dir23}
\begin{align}
i\hbar\partial^\lambda \ps = -\hbar \sigma^{\lambda \mu}\partial_\mu \ps+m \gamma^\lambda\ps, \\
-i \hbar\partial^\lambda \psbar=-\hbar\partial_\mu\psbar \sigma^{\lambda\mu} +m \psbar \gamma^\lambda.
\end{align}
\end{subequations}
Using \eqs\eqref{dir23} one can easily derive the Gordon decomposition~\cite{gordon1928strom}
\begin{equation}
\label{gordonn}
\psbar \gamma^\mu \ps = \frac{i \hbar}{2m}\left[ \psbar \olra{\partial}^\mu \ps -i \left( \psbar \sigma^{\mu\nu}\partial_\nu \ps -\partial_\nu \psbar \sigma^{\mu\nu} \ps\right) \right] .
\end{equation}
With the help of \eqs\eqref{dir23} and \eqref{gordonn}, the HW currents in \eqs\eqref{thw} and \eqref{shw} become
\begin{subequations}
\label{hw22}
\begin{align}
\T^{\mu\nu}_{HW}={}& \T^{\mu\nu}_{C} +\frac{i \hbar^2}{2m} (\partial^{\nu}\psbar \sigma^{\mu\beta}\partial_\beta \ps +\partial_\alpha \psbar \sigma^{\alpha\mu}\partial^{\nu}\ps) \n\\
&- \frac{i \hbar^2}{4m}g^{\mu\nu} \partial_\lambda (\psbar\sigma^{\lambda\alpha}\olra{\partial}_\alpha \ps) , \\
\Sp_{HW}^{\lambda,\mu\nu}={}&\Sp_{C}^{\lambda,\mu\nu}-\frac{\hbar^2}{4m}\left( \psbar \sigma^{\mu\nu} \sigma^{\lambda\rho}\partial_\rho \ps + \partial_\rho \psbar \sigma^{\lambda\rho} \sigma^{\mu\nu}\ps\right) .
\label{shw22}
\end{align}
\end{subequations}
Equations~\eqref{hw22} make apparent the connection between the canonical and the HW spin tensor. In the language of relocalization, Eqs.~\eqref{hw22} tell us that the HW currents can be obtained from the canonical ones through a pseudo-gauge transformation with the superpotentials~\cite{Hehl:1997ep,Kirsch:2001gt}
\begin{align}
  \hat{\Phi}^{\lambda,\mu\nu}&= \hat{M}^{[\mu\nu]\lambda}-g^{\lambda[\mu} \hat{M}_\rho^{\ \nu]\rho}, \label{pghw1}\\
   \hat{Z}^{\mu\nu\lambda\rho}&=-\frac{\hbar}{8m}\psbar(\sigma^{\mu\nu}\sigma^{\lambda\rho}+\sigma^{\lambda\rho}\sigma^{\mu\nu})\ps , \label{pghw2}
\end{align}
where
\begin{equation}
 \hat{M}^{\lambda\mu\nu}\equiv \frac{i\hbar^2}{4m}\psbar\sigma^{\mu\nu}\olra{\partial}^\lambda \ps .
\end{equation}

Since $\Sp_{HW}^{\lambda,\mu\nu}$ is conserved, the global HW spin defined from \eq\eqref{shw22} as
\begin{align}
\hat{S}^{\mu\nu}_{HW} &\equiv \int_\Sigma d\Sigma_\lambda\, \Sp_{HW}^{\lambda,\mu\nu} = \int d^3x\, \Sp_{HW}^{0,\mu\nu} \n\\
& = \frac\hbar 2\int d^3 x \, \ps^\dagger  \sigma^{\mu\nu} \ps + \frac{\hbar^2}{2m}\int d^3x\, \ps^\dagger \gamma^{[\mu} \partial^{\nu]} \ps ,
\label{stothw}
\end{align}
is a tensor, which is what we were searching for. In Eq. \eqref{stothw} we used Eq. \eqref{totcanspin} {and chose the hyperplane at constant $x^0$.}

\subsection{de Groot-van Leeuwen-van Weert currents}
\label{GLWsec}

Here we discuss another pair of currents which leads to the {same global spin as in the HW formulation, but different energy-momentum and spin tensors.} These currents are derived by de Groot, van Leeuwen and van Weert (GLW) in Ref.~\cite{DeGroot:1980dk} by performing the pseudo-gauge transformation
 \begin{eqnarray}
  \hat{\Phi}^{\lambda,\mu\nu}&=&\frac{i\hbar^2}{4m}\bar{\psi}(\sigma^{\lambda\mu} \overleftrightarrow{\partial}^\nu-\sigma^{\lambda\nu}\overleftrightarrow{\partial}^\mu)\psi,\\
  \hat{Z}^{\mu\nu\lambda\rho}&=&0.
 \end{eqnarray}
Thus, we obtain
\begin{align}
\label{tglw}
 \T^{\mu\nu}_{GLW}={}&-\frac{\hbar^2}{4m}\bar{\psi}\olra{\partial}^\mu\olra{\partial}^\nu\psi ,\\
 \Sp^{\lambda,\mu\nu}_{GLW}={}&\frac{i\hbar^2}{4m}\left( \bar{\psi}\sigma^{\mu\nu}\olra{\partial}^\lambda\psi-\partial_\rho \epsilon^{\mu\nu\lambda\rho}\psbar \gamma^5\ps \right),
 \label{sglw}
\end{align}
where we used the Gordon decomposition \eqref{gordonn}.
We note that \eq\eqref{sglw} differs from \eq\eqref{shw} only by a total divergence, hence for the global spin we have $\Sp^{\mu\nu}_{GLW}=\Sp^{\mu\nu}_{HW}$.

\subsection{Alternative Klein-Gordon currents}
\label{KGsec}

There is a possible choice of currents where the energy-momentum tensor is the same as in the GLW case and the spin tensor is the same as in the HW formulation.
To find such currents we follow the same idea as HW, but we now consider the Klein-Gordon Lagrangian for spinors built with second-order derivatives of the fields 
\begin{equation}
 \mathcal{L}^\prime_{KG}=\frac{1}{2m}\left\{-\frac{\hbar^2}{2}\left[(\partial_\mu\partial^\mu\bar{\psi})\psi+\bar{\psi}\partial_\mu\partial^\mu\psi\right]-m^2\bar{\psi}\psi\right\} .
\end{equation} 
Since $\mathcal{L}^\prime_{KG}$ and $\mathcal{L}_{KG}$ in \eq\eqref{lkg} differ only by a total divergence, they yield the same equations of motion, namely \eqs\eqref{kgeqs}. If we apply Noether's theorem using $\mathcal{L}^\prime_{KG}$, we obtain the following energy-momentum  and spin tensors~\footnote{For a Lagrangian with second-order derivatives of spinorial fields, the energy-momentum and spin tensors derived by applying Noether's theorem are given by~\cite{DeGroot:1980dk}
\begin{align*}
 T^{\mu\nu}={}&\frac{\partial\mathcal{L}^\prime_{KG}}{\partial(\partial_\mu\psi)} \partial^\nu\psi+( \partial^\nu\bar{\psi})\frac{\partial\mathcal{L}^\prime_{KG}}{\partial(\partial_\mu\bar{\psi})} \n\\
& \!\!\!\!\!\!\!\!\!\!\!\!\!\!\!+\frac{\partial\mathcal{L}^\prime_{KG}}{\partial(\partial_\rho\partial_\mu\psi)}\olra{\partial}_\rho\partial^\nu\psi - ( \partial^\nu\bar{\psi})\olra{\partial}_\rho\frac{\partial\mathcal{L}^\prime_{KG}}{\partial(\partial_\rho\partial_\mu\bar{\psi})}-g^{\mu\nu}\mathcal{L}^\prime_{KG},\\
  S^{\lambda,\mu\nu}={}&-\frac i{2}\left[\frac{\partial\mathcal{L}}{\partial(\partial_\lambda\psi)}\sigma^{\mu\nu}\psi-\bar{\psi}\sigma^{\mu\nu}\frac{\partial\mathcal{L}}{\partial(\partial_\lambda\bar{\psi})}\right. \n\\
 & \left.+\frac{\partial\mathcal{L}}{\partial(\partial_\rho\partial_\lambda\psi)}\olra{\partial}_\rho\sigma^{\mu\nu}\psi+\bar{\psi}\olra{\partial}_\rho\sigma^{\mu\nu}\frac{\partial\mathcal{L}}{\partial(\partial_\rho\partial_\lambda\bar{\psi})}\right].
\end{align*}
}
\begin{subequations} \label{kgcur}
\begin{align}
\label{tkg}
 \T^{\mu\nu}_{KG}={}&-\frac{\hbar^2}{4m}\bar{\psi}\olra{\partial}^\mu\olra{\partial}^\nu\psi -g^{\mu\nu}\mathcal{L}^\prime_{KG},\\
 \Sp^{\lambda,\mu\nu}_{KG}={}&\frac{i\hbar^2}{4m}\bar{\psi}\sigma^{\mu\nu}\olra{\partial}^\lambda\psi.
 \label{skg}
\end{align}
\end{subequations}
Since using the equations of motion $\mathcal{L}_{KG}^\prime=0$, we have $\T^{\mu\nu}_{KG}=\T^{\mu\nu}_{GLW}$, as can be seen from \eq\eqref{tglw}. Furthermore, from \eq\eqref{shw}, we see that $\Sp^{\lambda,\mu\nu}_{KG}=\Sp^{\lambda,\mu\nu}_{HW}$. 
The pseudo-gauge transformation to obtain \eqs\eqref{kgcur} is given by
\begin{eqnarray}
  \hat{\Phi}^{\lambda,\mu\nu}&=&\frac{i\hbar^2}{4m}\bar{\psi}(\sigma^{\lambda\mu} \overleftrightarrow{\partial}^\nu-\sigma^{\lambda\nu}\overleftrightarrow{\partial}^\mu)\psi,\\
  \hat{Z}^{\mu\nu\lambda\rho}&=& \frac{i\hbar^2}{4m}\epsilon^{\mu\nu\lambda\rho} \psbar \gamma^5 \ps.
 \end{eqnarray}

\section{Spin vector}
\label{pl_sec}

In nonrelativistic quantum mechanics, the spin vector operator $\hat{\bf S}_{nr}$ in first quantization  is simply given by 
\begin{equation}
\label{nrspin}
\hat{\bf S}_{nr}=\frac\hbar 2\boldsymbol{\sigma},
\end{equation}
where $\boldsymbol{\sigma}=(\sigma_1,\sigma_2,\sigma_3)$ are the Pauli matrices~\cite{Leader:2001}. 
In this section we examine two approaches to generalize the definition above to a covariant expression in quantum field theory: the first one is based on the Frenkel theory and the second one on the Pauli-Lubanski vector. {Furthermore, we show how the spin vectors in these approaches can be related to the various spin tensors introduced in Sec. \ref{secvarious}.
}

\subsection{Frenkel theory}
\label{secfrenk}

Following the idea of Frenkel~\cite{frenkel1926elektrodynamik} we
introduce an antisymmetric tensor  $\hat{S}^{\mu\nu}$ which depends on the total four-momentum $\hat{P}^\mu$ in \eq\eqref{totp}. 
In the particle rest frame, the spatial components of the four-momentum vanish, i.e., $\hat{P}^{i}\ket{\ppm_\star,s}=0$, where $\ppm_\star^\mu=(m,{\bf 0})$ is the momentum of the particle at rest. 
Furthermore, the components of $\hat{S}^{\mu\nu}$ are such that
\begin{align}
\bra{\ppm_\star,s}\hat{S}^{0i} \ket{\ppm_\star,s}&=0, \\
\bra{\ppm_\star,s}\hat{S}^{ij}\ket{\ppm_\star,s}&\equiv  \epsilon^{ijk}\bra{\ppm_\star,s}\hat{S}^k_{nr}\ket{\ppm_\star,s},
\label{spinrfcan}
\end{align}
where the operator on the right-hand side of \eq\eqref{spinrfcan} is also expressed in second quantization.
Thus, we establish a relation between the components of $\hat{S}^{\mu\nu}$ in the particle rest frame and those of the nonrelativistic spin vector \eqref{nrspin}.
In a compact form, for a particle state in a general frame, we write
\begin{equation}
\label{frenk}
\bra{\ppm,s}\hat{P}_{\mu}\hat{S}^{\mu\nu}\ket{\ppm,s}=0.
\end{equation}
The equation above is called the Frenkel condition.

Now we want to relate the Frenkel theory to the  pseudo-gauges discussed in the previous sections.
In particular, we want to find for which pseudo-gauge choice the tensor $\hat{S}^{\mu\nu}$ introduced in this section can be identified with the global spin. 
Consider first the canonical global spin in \eq\eqref{totcanspin}. We note that, $\hat{S}_C^{0i}=0$ and
\begin{align}
 \hat{S}_C^{ij}&\equiv  \epsilon^{ijk}\hat{S}_C^k,
\end{align}
with 
\begin{equation}
\label{canspinv}
\hat{S}^k_C = \int d^3x\,  \ps^\dagger \frac\hbar 2 {\mathfrak{S}}^k\ps,
\end{equation}
where the integration over the hypersurface is performed choosing a hyperplane with constant  $x^0$, and
\begin{equation}
\label{bolds}
\mathfrak{S}^k =
\begin{pmatrix}
\sigma^k & 0 \\
0 & \sigma^k 
\end{pmatrix} .
\end{equation}
If we take the expectation value of a one-particle state at rest, \eq\eqref{canspinv} is the expression in second quantization of the nonrelativistic spin operator \eqref{nrspin}. We see again that $\hat{S}_C^{\mu\nu}$ is not a tensor and it is clearly not compatible with the Frenkel condition since, in this case, \eq\eqref{frenk} holds only when taking the expectation value of a state of particle at rest.

Let us now turn to the HW (or equivalently the GLW or KG) global spin \footnote{In the following,
since  $\Sp^{\mu\nu}_{HW}=\Sp^{\mu\nu}_{GLW}=\Sp^{\mu\nu}_{KG}$, we will only use the HW subscript for the sake of ease of notation.} in \eq\eqref{stothw}. In a general frame the components $\hat{S}_{HW}^{0i}$ do not vanish and
\begin{equation}
\hat{S}_{HW}^{ij}=  \epsilon^{ijk}\hat{S}_{HW}^k
\end{equation}
with 
\begin{equation}
\label{shwgf}
\hat{S}_{HW}^k=\int d^3 x\, \psi^\dagger\left(\frac\hbar 2\mathfrak{S}^k+\frac{\hbar^2}{ 2m}\epsilon^{kln}\gamma^l {\partial}^n \right)\psi .
\end{equation}
One can check by an explicit calculation using similar steps as those outlined in App. \ref{apppp2} that, for a particle at rest, 
\begin{equation}
\bra{\ppm_\star,s}\hat{S}_{ HW}^{0i}\ket{\ppm_\star,s}=0,
\end{equation}
the second addend in \eq\eqref{shwgf} vanishes and, from \eq\eqref{canspinv}, we see that 
\begin{equation}
\label{hwcc}
\bra{\ppm_\star,s}\hat{S}_{HW}^{ij}\ket{\ppm_\star,s}=\bra{\ppm_\star,s}\hat{S}_{C}^{ij}\ket{\ppm_\star,s} .
\end{equation} 
This means that the space components of the HW global spin reduce to the nonrelativistic spin vector and  that  $\hat{S}^{\mu\nu}_{HW}$  behaves as a tensor in accordance with the Frenkel theory \eqref{frenk}, i.e.,
\begin{equation}
\label{frenkelhwhw}
\bra{\ppm,s}\hat{P}_\mu\hat{S}_{HW}^{\mu\nu}\ket{\ppm,s}=0.
\end{equation}
Thus, the HW global spin gives a covariant generalization of the nonrelativistic spin operator.

\subsection{Pauli-Lubanski vector}
\label{plvec}

In this subsection, we generalize the nonrelativistic spin operator in \eq\eqref{nrspin} to a covariant  vector $\hat{S}^\mu$ (even though strictly speaking it is a pseudovector, we will use the term vector for simplicity). This vector is such that, in the particle rest frame, it reduces to the form
\begin{equation}
\bra{\ppm_\star,s}\hat{S}^\mu \ket{\ppm_\star,s}=(0,\bra{\ppm_\star,s}\hat{\bf S}_{nr}\ket{\ppm_\star,s})
\end{equation}
or, covariantly,
\begin{equation}
\bra{\ppm,s}\hat{P}_\mu \hat{S}^{\mu}\ket{\ppm,s}=0.
\end{equation}
In order to define the relativistic spin operator, we introduce the Pauli-Lubanski vector~\cite{itzykson2012quantum,Leader:2001} 
\begin{equation}
\label{w1233}
\hat{w}^\mu = -\frac12 \epsilon^{\mu\nu\alpha\beta}\hat{P}_\nu \hat{J}_{\alpha\beta}
\end{equation}
where $\hat{J}_{\mu\nu}$ is the total angular momentum in \eq\eqref{totj}. Using the commutation relations of the Poincar{\'e} algebra we obtain~\cite{Leader:2001}
\begin{equation}
\label{comm0}
[\hat{w}^\mu , \hat{w}^\nu]=-i\hbar\,\epsilon^{\mu\nu\alpha\beta} \hat{w}_\alpha \hat{P}_\beta .
\end{equation}
If we consider the action of the commutator on states at rest, then for the spatial components we have
\begin{equation}
[\hat{w}^i , \hat{w}^j]=-i\hbar\,\epsilon^{ijk0} \hat{w}_k m .
\end{equation}
Therefore, the relativistic spin operator can be defined as
\begin{equation}
\label{spsp}
\hat{S}^\mu =\frac{\hat{w}^\mu}{ m},
\end{equation}
since its spatial components follow the usual commutation relations for spin operators
\begin{equation}
\label{comm}
[\hat{S}_i , \hat{S}_j] = i\hbar\, \epsilon_{ijk} \hat{S}_k,
\end{equation}
provided that they act onto states at rest.

We note that, since the total charges are pseudo-gauge invariant by construction, it follows that the relativistic spin vector is also a pseudo-gauge invariant quantity. {Inserting \eq\eqref{totj} in \eqs\eqref{w1233} and \eqref{spsp}, and taking the matrix element of one-particle states, one obtains
\begin{align}
&\bra{ \ppm^\prime,s^\prime} \hat{S}^\mu \ket{\ppm,s} \n\\ 
={}&-\frac1{2m}\epsilon^{\mu\nu\alpha\beta}\bra{ \ppm^\prime,s^\prime}\hat{P}_\nu\Sp_{C,\alpha\beta}\ket{\ppm,s}\n\\
={}&-\frac1{2m}\epsilon^{\mu\nu\alpha\beta}\bra{ \ppm^\prime,s^\prime}\hat{P}_\nu\Sp_{HW,\alpha\beta}\ket{\ppm,s} \n\\
={}&-\frac1{2m}\epsilon^{\mu\nu\alpha\beta}\ppm_\nu\int d^3x\, \bra{ \ppm^\prime,s^\prime} \ps^\dagger(x) \frac\hbar2 \sigma_{\alpha\beta} \ps(x)\ket{\ppm,s},
\label{whatweprove0}
\end{align}
for the details of the derivation see App. \ref{apppp2}.
The equation above shows that the contribution of the orbital parts of the canonical and HW pseudo-gauge vanish when one-particle states are considered. 
 For a related discussion, see Refs.~\cite{Becattini:2020sww,Tinti:2020gyh}.
 Finally, we note that the inverse relation involving the expectation value of one-particle states of the form
\begin{equation}
 \bra{ \ppm,s}\hat{S}_{HW}^{\mu\nu}\ket{\ppm,s}=-\frac{1}{m}\epsilon^{\mu\nu\alpha\beta} \bra{ \ppm,s}\hat{P}_\alpha\hat{S}_\beta \ket{\ppm,s}
\end{equation}
is only valid for the HW, GLW and KG (and not for the canonical) pseudo-gauge since in this case \eq\eqref{frenkelhwhw} holds.
We can thus write
\begin{equation}
\bra{ \ppm,s}\hat{S}^{\mu\nu}_v \ket{\ppm,s}=-\epsilon^{\mu\nu\alpha\beta}  \bra{ \ppm,s} v_\alpha\hat{S}_\beta \ket{\ppm,s},
\end{equation}
where $\hat{S}^{\mu\nu}_v=\hat{S}^{\mu\nu}_C$ for ${v}^\mu=(1,{\bf 0})$ and $\hat{S}^{\mu\nu}_v=\hat{S}^{\mu\nu}_{HW}$ for ${v}^\mu=\hat{P}^\mu/m$.
}

We conclude this section with a brief remark about massless particles. It is clear that the spin vector in Eq. \eqref{spsp} is not defined for particles with vanishing mass. We note that the action of the commutator of the Pauli-Lubanski vectors \eqref{comm0} on a massless one-particle state  cannot be reduced to the conventional commutation relations for spin operators \eqref{comm}, since the physical quantum number is now the particle helicity~\cite{Leader:2001}. {It is possible to show that the action of the Pauli-Lubanski vector on a massless one-particle state $| \mathfrak{p}, \lambda \rangle$ with helicity $\lambda=\pm1/2$, is given by \cite{Leader:2001,atre1987massless}
\begin{equation}
\label{helicity}
\hat{w}^\mu \ket{ \mathfrak{p}, \lambda }= \hbar \lambda \hat{P}^\mu \ket{ \mathfrak{p}, \lambda }.
\end{equation}
Equation \eqref{helicity}
shows that for massless particles the spin is slaved to the momentum.
}

\section{Wigner operator}
\label{wigsec}

Quantum mechanics can be equivalently formulated with the so-called Wigner function~\cite{Wigner:1932eb}, namely a distribution function in phase-space which can be alternatively used to calculate expectation values of operators. In this sense, the Wigner function generalizes the concept of classical distribution function to the quantum case. However, caution must be taken when making this identification, since the Wigner function can in general take on complex values. 
In quantum field theory the Wigner operator for the Dirac field is defined as~\cite{DeGroot:1980dk,Heinz:1983nx,Elze:1986qd,Vasak:1987um}
 \begin{equation}
 \label{wtrans}
 \hat{ W}_{\kappa\chi}(x,p)=\int \frac{d^4y}{(2\pi\hbar)^4} e^{-\frac i\hbar p\cdot y} \psbar_\chi\left(x_1\right)\ps_\kappa\left(x_2\right),
 \end{equation}
with $x_1=x+y/2$, $x_2=x-y/2$ and $\kappa, \ \chi$ denote here Dirac indices. Using the Dirac equation for free fields \eqref{diraceq1} and \eqref{diraceq2},
one derives the equation of motion for the Wigner operator \cite{DeGroot:1980dk,Elze:1986qd,Vasak:1987um}
\begin{equation}
 \left[ \gamma \cdot \left( p+i \,\frac{\hbar}{2} \partial \right) -m\right] \hat{W}(x,p)=0 , \label{Wignerkin}
\end{equation}
Applying the operator $[\gamma \cdot \left( p+i \,\frac{\hbar}{2} \partial \right) +m]$ to \eq\eqref{Wignerkin} and separating real and imaginary part we obtain
\begin{align}
\left( p^2 - m^2 -\frac{\hbar^2}{4}\partial^2\right) \hat{W}(x,p) &=0, \label{eqwig11}\\
p\cdot \partial\, \hat{W}(x,p) &=0, \label{eqwig22}
\end{align}
respectively.
We recognize in \eq\eqref{eqwig11} the modification of the on-shell condition defining the particle spectrum and in \eq\eqref{eqwig22} a Boltzmann-like equation.

It is convenient to decompose the Wigner operator in terms of a basis of the generators of the Clifford algebra 
\begin{equation}
\hat{W}=\frac14\left(\hat{\F}+i\gamma^5\hat{\Pc}+\gamma^\mu \hat{\V}_\mu+\gamma^5\gamma^\mu \hat{\A}_\mu
+\frac12\sigma^{\mu\nu}\hat{\Sc}_{\mu\nu}\right), \label{clifdec}
\end{equation}
with the coefficients given by
\begin{subequations}
\label{coef0}
\begin{align}
\hat{\F}={}&\Tr(\hat{W}),  \\
\hat{\Pc}={}&-i\,\Tr(\gamma^5\hat{W}),\\
\hat{\V}^\mu={}&\Tr(\gamma^\mu \hat{W}),\\
\hat{\A}^\mu={}&\Tr(\gamma^\mu\gamma^5 \hat{W}),\\
\hat{\Sc}^{\mu\nu}={}&\Tr(\sigma^{\mu\nu} \hat{W}), 
\end{align}
\end{subequations}
where the traces are meant over the Dirac indices. We substitute \eq\eqref{clifdec} into \eq\eqref{Wignerkin} and decompose real and imaginary part to obtain the equations of motion for the coefficient functions. We write here only the two equations we will use in this section (for the complete set of equations of motion for the coefficients when also interactions are considered see \secc\ref{secspinhyd}) 
\begin{subequations}
\begin{align}
p\cdot \hat{\V} -m\hat{\F}&= 0, \label{plag} \\
p^\mu \hat{\F} - \frac\hbar 2 \partial_\nu \hat{\Sc}^{\nu\mu}-m\hat{\V}^\mu&=0, \label{vvvvvv}\\
-\frac\hbar 2 \partial^\mu \hat{\Pc} +\frac12 \epsilon^{\mu\nu\alpha\beta}p_\nu \hat{\Sc}_{\alpha\beta} + m \hat{\A}^\mu &=0, \label{aaaaaaaa} \\
p \cdot \hat{\A}&=0,\label{orth}\\
\frac{\hbar}{2}\partial^\mu \hat{\F}+p_\nu \hat{\Sc}^{\nu\mu}&=0.\label{B}
\end{align}
\end{subequations}
The energy-momentum and spin tensors discussed in \secc\ref{secvarious} can be easily expressed in terms of the functions \eqref{coef0}. In particular, \eqs\eqref{tcan}, \eqref{tbel}, \eqref{thw} and \eqref{tglw} can be written as
\begin{align}
 \T^{\mu\nu}_{C}&=\int d^4 p\,  p^{\nu} \hat{\V}^{\mu} , \label{tcanw}\\
 \T^{\mu\nu}_{B}&=\frac12 \int d^4p\, (p^\nu \hat{\V}^\mu+p^\mu \hat{\V}^\nu), \label{TBBB}\\
 \T^{\mu\nu}_{HW}&= \frac1m\int d^4p\, \left[p^\mu p^\nu+\frac{\hbar^2}{4}\left(\partial^\mu\partial^\nu-g^{\mu\nu}\partial^2\right)\right]\hat{\F}, \label{thww} \\
 \T^{\mu\nu}_{GLW}&=\T^{\mu\nu}_{KG}=\frac1m \int d^4p\, p^\mu p^\nu \hat{\F} , \label{tglww}
\end{align}
respectively. {In the equations above, the free equation of motion \eqref{plag} and the on-shell condition \eqref{eqwig11} were used to eliminate the term proportional to $g^{\mu\nu}$ with the Dirac and the Klein-Gordon Lagrangian in the energy-momentum tensors \eqref{tcanw}, \eqref{TBBB} and \eqref{tglww}. We note that for the HW energy-momentum tensor, the term proportional to $g^{\mu\nu}$ cannot be eliminated with the equations of motion.} The spin tensors in \eqs\eqref{scan}, \eqref{shw}, \eqref{sglw} can be now expressed as
\begin{align}
 \Sp^{\lambda,\mu\nu}_{C}&=-\frac\hbar 2\epsilon^{\lambda\mu\nu\rho}\int d^4p\, \hat{\A}_\rho, \label{scanw}\\
% \Sp^{\lambda,\mu\nu}_{Bel}&=\frac12 x^{[\mu}\int d^4p\, \left(p^{\nu]}\V^\lambda-p^{\lambda]}\V^\nu\right),\\
 \Sp^{\lambda,\mu\nu}_{HW}&= \Sp^{\lambda,\mu\nu}_{KG}=\frac{\hbar}{2m}\int d^4p\, p^\lambda \hat{\Sc}^{\mu\nu}, \label{spinwigner}\\
\Sp^{\lambda,\mu\nu}_{GLW}&=\frac{\hbar}{2m}\int d^4p\,\left( p^\lambda \hat{\Sc}^{\mu\nu} -\frac\hbar2 \epsilon^{\lambda\mu\nu\alpha}\partial_\alpha \hat{\Pc}\right).
\label{spinglwtrue}
\end{align}

We now want to relate the spin vector to the Wigner operator. In accordance with Ref. \cite{Becattini:2020sww}, we define the spin vector as
\begin{equation}
 \hat{\Pi}^\mu(p)= -\frac{1}{2m}\epsilon^{\mu\nu\alpha\beta}p_\nu \hat{j}_{\alpha\beta}(p), \label{pltat}
 \end{equation}
where $\hat{j}_{\alpha\beta}$ is related to the total angular momentum in \eq\eqref{totj} through $\hat{J}_{\alpha\beta}=\int d^4p\, \hat{j}_{\alpha\beta}$. 
Let us justify Eq. \eqref{pltat} using the Frenkel theory. To do so we define the quantity 
\begin{align}
\hat{s}^{\mu\nu}_{HW}(p)&=\frac \hbar {2m}\int_\Sigma d\Sigma_\lambda \, p^\lambda \hat{\Sc}^{\mu\nu}(x,p) \n\\ 
&=\hbar \frac{p^0}{2m} \int d^3x  \, \hat{\Sc}^{\mu\nu}(x,p).
\end{align}
{It is easy to check that $\hat{s}^{\mu\nu}_{HW}$ is a Lorentz tensor, since $p^\lambda\partial_\lambda \hat{\Sc}^{\mu\nu}=0$ which, in turn, is a consequence of  \eq\eqref{eqwig22}.} The global spin can be written as
\begin{equation}
\Sp^{\mu\nu}_{HW} = \int d^4 p \, \hat{s}^{\mu\nu}_{HW}(p).
\end{equation}
From \eq\eqref{B} we get
\begin{equation}
 p_\mu \hat{s}^{\mu i}_{HW}=-\frac{\hbar^2}{2m}p^0\int d^3x\, \partial^i \hat{\F}=0,
\end{equation}
where boundary terms were neglected, and 
\begin{equation}
 p_\mu \hat{s}^{\mu0}_{HW}=-\frac{\hbar^2}{2m}\int d^3x\, p\cdot\partial \hat{\F}=0,
\end{equation}
which follows again from \eq\eqref{eqwig22}. Therefore, we deduce that 
\begin{equation}
\label{frenk2}
p_\mu \hat{s}^{\mu i}_{HW} = 0 .
\end{equation}
Equation~\eqref{frenk2} shows that, for free fields, $\hat{s}^{\mu \nu}_{HW}$ satisfies a form of the Frenkel condition. The difference with \eq\eqref{frenk} is that here $p^\mu$ is the conjugate variable of $y^\mu	$ in the Wigner transform \eqref{wtrans} and not in general the total momentum of the system (or of the particle). In other words the spin in the rest frame of the momentum operator is equivalent to defining the spin in the rest frame of the momentum variable $p^\mu$. Therefore, we can define the  spin vector based on the variable $p^\mu$ and the  HW spin tensor \eqref{spinwigner} as
\begin{align}
 \hat{\Pi}^\mu(p)&\equiv -\frac{1}{2m}\epsilon^{\mu\nu\alpha\beta}p_\nu \hat{s}_{HW,\alpha\beta}(p) \n\\
 &= -\frac{\hbar}{4m^2}\epsilon^{\mu\nu\alpha\beta}p_\nu \int_\Sigma d\Sigma^\lambda \, p_\lambda \hat{\Sc}_{\alpha\beta} (x,p) \n\\
&= -\hbar\frac{p^0}{4m^2}\epsilon^{\mu\nu\alpha\beta}p_\nu \int d^3x\, \hat{\Sc}_{\alpha\beta} (x,p),
 \label{plpl2}
\end{align}
where in the last line we chose the hypersurface integration at constant $x^0$. We stress that, although expressed in terms of the HW spin tensor, Eq. \eqref{plpl2} is pseudo-gauge invariant and is equal to  Eq. \eqref{pltat}. To show this, we only need to prove that the HW orbital part  of the spin vector obtained using Eq. \eqref{thww},
\begin{equation}
\label{asd123322}
%&-\frac{1}{2m}\epsilon^{\mu\nu\alpha\beta}p_\nu \int d\Sigma^\lambda x_{[\alpha} \hat{t}_{HW,\beta]\lambda}(p) \n \\
-\frac{1}{2m^2}\epsilon^{\mu\nu\alpha\beta}p_\nu \left[ \int d^3 x\, x_{[\alpha}  p_{\beta]} p_0 \hat{\F} + \frac{\hbar^2}{4} l_{\alpha\beta} \right]  ,
\end{equation}
vanishes, where
\begin{equation}
l_{\alpha\beta} = \int d^3 x\, x_{[\alpha} \left( \partial_{\beta]}\partial_0 -  g_{\beta ] 0}\partial^2\right)\hat{\F} .
\end{equation}
Note that both terms in Eq. \eqref{asd123322} inside the square brackets are tensors. The first one vanishes when contracted with the Levi-Civita tensor. For the second one, we have $l_{\alpha\beta}=0$, which is what must hold, since the HW and GLW pseudo-gauges have the same global orbital angular momentum [compare, e.g., Eqs. \eqref{thww} and \eqref{tglww}]. To see this, let us consider each component of $l_{\alpha\beta}$ separately. 
For $\alpha=0$, $\beta=i$, we have
\begin{align}
l_{0i}={}&\int d^3 x\, (x_{0}\partial_0\partial_{i}-x_i \partial_0^2 +x_i\partial^2)\hat{\F}] \n\\
={}& x_{0}\partial_0\int d^3 x\,  \partial_{i}\hat{\F} + \int d^3 x\, x_i \partial_j \partial^j \hat{\F} \n\\
={}&g_{ij}\int  d^3 x\,  \partial^j \hat{\F} =0,
\end{align}
where we integrated by parts the second integral in the second line and neglected boundary terms. For $\alpha=i$, $\beta=j$ we have
\begin{equation}
l_{ij}=\partial_0\int d^3 x\, (x_{i} \partial_{j}-x_{j} \partial_{i})\hat{\F} = 0,
\end{equation}
which follows after again integration by parts of both terms.

{Using the canonical currents \eqref{tcanw} and \eqref{scanw}, \eq\eqref{pltat} can also be written as
\begin{align}
 \hat{\Pi}^\mu(p)&= -\frac{1}{2m}\epsilon^{\mu\nu\alpha\beta}p_\nu \hat{s}_{C,\alpha\beta}(p) \n\\
 &=\frac{\hbar}{2m}\int_\Sigma d\Sigma_\lambda \, p^\lambda \hat{\A}^\mu  (x,p) \n\\
 &=\hbar\frac{p^0}{2m} \int d^3\, x \, \hat{\A}^\mu (x,p),
  \label{plpl3}
\end{align}
where from the first to the second step we used \eq\eqref{orth}. It is possible to show that Eqs. \eqref{plpl3} and \eqref{plpl2} are equal by using the equation of motion \eqref{aaaaaaaa} and the Boltzmann equation $p\cdot\partial \hat{\Pc}=0$ which follows from Eq. \eqref{eqwig22}.} Note that, however, the inverse relation 
\begin{equation}
 \hat{s}_{HW}^{\mu\nu}(p)=-\frac1m\epsilon^{\mu\nu\alpha\beta}p_\alpha \hat{\Pi}_\beta (p)
\end{equation}
is only valid for the HW, GLW and KG (but not canonical) spin tensor. {Finally, if we use the Belinfante pseudo-gauge in Eq. \eqref{pltat}, the spin vector has only the contribution of the orbital part of the angular momentum, since the spin tensor vanishes. Thus, using the Belinfante energy-momentum tensor \eqref{TBBB}, we can express the spin vector \eqref{pltat} as
\begin{align}
 \hat{\Pi}^\mu(p)&= -\frac{1}{4m}\epsilon^{\mu\nu\alpha\beta}p_\nu \int_\Sigma d\Sigma^\lambda p_\lambda x_{[\alpha} \hat{\V}_{\beta]} (x,p)  \n\\
 &=-\frac{p^0}{2m}\epsilon^{\mu\nu\alpha\beta}p_\nu \int d^3x\, x_{\alpha} \hat{\V}_{\beta} (x,p) ,
  \label{plpl4}
\end{align}
which can be shown to be equal to Eqs. \eqref{plpl2} and \eqref{plpl4} after employing Eq. \eqref{vvvvvv}, integrating by parts, and utilizing $p\cdot\partial \hat{\Sc}^{\mu\nu}=0$ which follows from Eq. \eqref{eqwig22}.}

As a concluding remark of this section, we mention that the operator $\hat{\Pi}^\mu (p)$ should not be confused with the spin vector defined in \secc\ref{pl_sec}. In fact, one can see that $\int d^4p\, \hat{\Pi}^\mu (p)$ is in general different from $\hat{S}^\mu$ in \eq\eqref{spsp}. 
However, when taking the matrix element on one-particle states, we obtain the relation
\begin{equation}
\label{seqpi}
\bra{ \ppm^\prime,s^\prime} \hat{S}^\mu\ket{\ppm,s} = \int d^4p\, \bra{ \ppm^\prime,s^\prime}\hat{\Pi}^\mu (p)\ket{\ppm,s} . 
\end{equation}
We emphasize that \eq\eqref{seqpi} is only valid for free fields. In this case one can prove that the momentum of the Wigner function is set on-shell by the spacetime integration \cite{Becattini:2020sww,Tinti:2020gyh}.
For more details see \app~\ref{apppp2}.

\section{Pseudo-gauges and statistical operator}
\label{nonsec}

So far, we have only dealt with operators and their matrix element of one-particle states.
A natural question we can now ask is how different pseudo-gauges affect the thermodynamic description of a system~\cite{Becattini:2011ev,Becattini:2012pp,Becattini:2018duy,Becattini:2020riu}. 
To address this question, in this section we study the consequences
of the pseudo-gauge transformations on  the statistical operator $\hat{\rho}$~\cite{Becattini:2018duy}.
In statistical quantum field theory a possible way to determine $\hat{\rho}$ is by using the method proposed by Zubarev~\cite{Zubarev1966} and later rediscussed in Refs.~\cite{van1982maximum,Becattini:2014yxa,Hongo:2016mqm,Becattini:2019dxo}. The local equilibrium density operator $\hat{\rho}_\text{LE}$ is obtained by maximizing the entropy $s=-\tr(\rr \log \rr)$ imposing  constraints on the energy-momentum and total angular momentum to be equal to the actual ones. Let us first start with the Belinfante case in which  the constraints are given by
\begin{align}
\nn_\mu \tr (\rr_B \, \wT^{\mu\nu}_B) &= \nn_\mu T^{\mu\nu}_B,  \label{cb1}\\
   \nn_\mu \tr ( \rr_B \, \hat{J}_B^{\mu,\lambda\nu} ) 
   &= \nn_\mu \tr [ \rr_B ( x^\lambda \wT_B^{\mu\nu} - 
    x^\nu \wT_B^{\mu\lambda}  ) ] 
    = \nn_\mu {J}_B^{\mu,\lambda\nu},
     \label{cb2}
\end{align}
where on the right-hand sides the quantities $T_B^{\mu\nu}$ and ${J}_B^{\mu,\lambda\nu}$ are the actual densities. The vector $\nn$ in \eqs\eqref{cb1} and \eqref{cb2} is the normal to some hypersurface $\Sigma$ that we defined by a proper foliation of the spacetime.
Notice that \eq\eqref{cb1} implies that \eq\eqref{cb2} is redundant: once we have the constraint on the energy-momentum tensor we automatically have that on the total angular momentum. Thus, the local equilibrium density operator reads
\begin{equation}
\label{rrb}
\rr_{B, \text{LE}} 
= \frac{1}{Z_B} \exp \left[-\int_\Sigma d \Sigma_\mu  \wT^{\mu\nu}_B 
\beta_{B,\nu}   \right],
\end{equation}
where $Z_B=\Tr \rr_{B, \text{LE}}$ and $\beta_{B,\nu}$ is the Lagrange multiplier associated to momentum conservation. We stress that the Lagrange multiplier depends on the choice of the pseudo-gauge because it has to be a solution of the constraint at local equilibrium
\begin{equation}
\nn_\mu \Tr \Big[\rr_{B, \text{LE}}(\beta_{B}) \, \wT^{\mu\nu}_B\Big] = \nn_\mu T^{\mu\nu}_B (\beta_{B}),
\end{equation}
which are four equations for the four unknowns $\beta_{B, \nu}$. The operator \eqref{rrb}, however, being time dependent, is not the real density operator in the Heisenberg picture. {The true statistical operator $\hat{\rho}_{\text{LE},0}$ is assumed to be the one in \eq\eqref{rrb} evaluated
 at some specific time with corresponding hypersurface $\Sigma_0$ where the system is known to be in local thermodynamic equilibrium \cite{Zubarev1966}.} The true statistical operator is what one needs for the calculation of the ensemble average of any operator $\hat{O}$, 
\begin{equation}
O\equiv \langle \hat{O} \rangle=\Tr (\hat{\rho}_{\text{LE},0}\,\hat{O}).
\end{equation}

Let us now follow the same steps to construct the density operator using the canonical currents. We immediately see that the constraint on the angular momentum is not redundant anymore because we have a nonvanishing spin tensor, i.e.,
\begin{align}
\nn_\mu \tr (\rr_C \, \wT^{\mu\nu}_C) &= \nn_\mu T^{\mu\nu}_C,  \label{cc1}\\
   \nn_\mu \tr ( \rr_C \, \hat{J}_C^{\mu,\lambda\nu} ) 
   &= \nn_\mu \tr [ \rr_C ( x^\lambda \wT_C^{\mu\nu} - 
    x^\nu \wT_C^{\mu\lambda}  + \hat{S}_C^{\mu,\lambda\nu}) ] \nonumber \label{cc2}\\
    &= \nn_\mu {J}_C^{\mu,\lambda\nu}.
\end{align}
Therefore, \eq\eqref{cc2} reduces to the effective independent constraint on the spin tensor
\begin{equation}
\nn_\mu \tr (\rr_C \, \hat{S}^{\mu,\lambda\nu}_C) = \nn_\mu S^{\mu,\lambda\nu}_C,
\end{equation}
and the canonical local equilibrium density operator is given by
\begin{equation}
\label{rrc}
\rr_{C,\rm LE} = \frac{1}{Z_{C}} \exp \left[-\int_\Sigma d \Sigma_\mu (\hat{T}_C^{\mu\nu} \beta_{C,\nu} 
 - \frac{1}{2} \hat{S}_C^{\mu,\lambda\nu}\Omega_{C,\lambda\nu}) \right],
\end{equation}
with $Z_{C}=\Tr \rr_{C, \text{LE}}$. The quantity $\Omega_{C}^{\lambda\nu}$ is called spin potential and corresponds to the Lagrange multiplier related to the conservation of the total angular momentum. The fields $\beta_C$ and $\Omega_C$ are solutions of the equations
\begin{align}
\nn_\mu \Tr \Big[\rr_{C, \text{LE}}(\beta_{C},\Omega_C) \, \wT^{\mu\nu}_C\Big] &= \nn_\mu T^{\mu\nu}_C (\beta_{C},\Omega_C), \\
\nn_\mu \Tr \Big[\rr_{C, \text{LE}}(\beta_{C},\Omega_C) \, \hat{S}_C^{\mu,\lambda\nu}\Big] &= \nn_\mu
{S}_C^{\mu,\lambda\nu} (\beta_{C},\Omega_C).
\end{align}

It is clear that, in general, \eqs\eqref{rrb} and \eqref{rrc} are not equal. In order to compare the two density operators, we perform in \eq\eqref{rrc} the pseudo-gauge transformation with the superpotentials in \eq\eqref{spbel} and we obtain 
\begin{align}
\label{rrbc}
&\rr_{C,\rm LE} = \frac{1}{Z_C} \exp \bigg[-\int_\Sigma d \Sigma_\mu \Big(\wT^{\mu\nu}_B\beta_{C,\nu}
  \nonumber\\
& - \frac{1}{2} (\Omega_{C,\lambda\nu}-\varpi_{C,\lambda\nu})\hat{S}_C^{\mu, \lambda \nu} + \frac{1}{2} \chi_{C,\lambda\nu} ( \hat{S}_C^{\lambda, \mu \nu} + \hat{S}_C^{\nu, \mu \lambda})\Big)\bigg],
\end{align}
where {we made use of the integration by parts and neglected boundary terms. Furthermore we defined}
\begin{align}
\label{thvort}
\varpi_C^{\lambda\nu} &=- \frac{1}{2} (\partial^\lambda \beta^{\nu}_C-\partial^\nu \beta^{\lambda}_C), \\
\chi_C^{\lambda\nu} &= \frac{1}{2} (\partial^\lambda \beta^{\nu}_C+\partial^\nu \beta^{\lambda}_C) .
\end{align}
The tensor in \eq\eqref{thvort} is called thermal vorticity. We note that, since the canonical spin tensor \eqref{scan} is antisymmetric under the exchange of all indices, then $\hat{S}_C^{\lambda, \mu \nu} + \hat{S}_C^{\nu, \mu \lambda}=0$ and the last term in the exponent in \eq\eqref{rrbc} vanishes. 
By comparing the two statistical operators \eqref{rrbc} and \eqref{rrc}, we infer that they are equal if  $\beta^\nu_{C}=\beta^\nu_{B}$, and
\begin{equation}
\label{ccoo11}
\Omega^{\lambda\nu}=\varpi^{\lambda\nu},
\end{equation}
where we removed the subscript $C$, since the $\beta^\nu$ fields in the different pseudo-gauges have to be the same.

{We can generalize what discussed so far by considering a generic pseudo-gauge transformation. If we start from the canonical statistical operator in Eq. \eqref{rrc} and perform the transformation in Eq.  \eqref{pgt}, we obtain
\begin{align}
\label{rrbc2}
&\rr_{C,\rm LE} = \frac{1}{Z_C} \exp \bigg[-\int_\Sigma d \Sigma_\mu \Big(\wT_\trans^{\mu\nu}\beta_{C,\nu} - \frac{1}{2} \hat{S}_\trans^{\mu,\lambda\nu}\Omega_{C,\lambda\nu}
 \nonumber\\
&  - \frac{1}{2} (\Omega_{C,\lambda\nu}-\varpi_{C,\lambda\nu})\hat{\Phi}^{\mu, \lambda \nu} + \frac{1}{2} \chi_{C,\lambda\nu} ( \hat{\Phi}^{\lambda, \mu \nu} + \hat{\Phi}^{\nu, \mu \lambda}) \nonumber \\
&-(\partial_\rho \Omega_{C,\lambda\nu}) \hat{Z}^{\lambda\nu,\mu\rho}
\Big)\bigg].
\end{align}
After comparing with the statistical operator constructed with the currents $\hat{T}_\trans^{\mu\nu}$ and $\hat{S}_\trans^{\mu,\lambda\nu}$,
\begin{align}
\label{rrc222}
\rr_{\trans , \rm LE} = {}&\frac{1}{Z_\trans} \exp \bigg[-\int_\Sigma d \Sigma_\mu \Big(\hat{T}_\trans^{\mu\nu} \beta_{\trans ,\nu}  \n\\
 {}&- \frac{1}{2} \hat{S}_\trans^{\mu,\lambda\nu}\Omega_{\trans,\lambda\nu}\Big) \bigg],
\end{align}
where $Z_\trans=\Tr \rr_{\trans , \text{LE}}$, we can readily get the conditions for the equivalence of the states, namely: $\beta^\mu_C=\beta_\trans^{\mu}$, $\Omega_{C}^{\lambda\nu}=\Omega_\trans^{\lambda\nu}$, Eq. \eqref{ccoo11} has to hold, and
\begin{equation}
\label{cond123321}
\chi^{\lambda\nu}=0, \qquad \partial^\rho \Omega^{\lambda\nu}=0.
\end{equation} 
Equation \eqref{ccoo11} together with Eq. \eqref{cond123321} are the conditions for global equilibrium \cite{Becattini:2012tc}. Therefore, we conclude that, in general, the local equilibrium statistical operator is a pseudo-gauge dependent quantity. Only in global equilibrium the statistical operators derived with different pseudo-gauges are equal, provided that the thermodynamic fields are the same. An important consequence of what showed in this section is that the expectation value of a generic operator $\langle \hat{O} \rangle$ will depend on the pseudo-gauge if the statistical operator describes a many-body state away from global equilibrium. Only in global equilibrium the expectation values calculated with different pseudo-gauges will be the same. Therefore, studying the expectation values of observables away from global equilibrium is in principle a possible way to understand which pseudo-gauge best describes the system.
}

\section{Spin-polarization effects in relativistic nuclear collisions}
\label{hicsec}

The formalism discussed so far is a powerful tool to study spin dynamics of relativistic many-body systems. In this section we will focus on applications to the physics of relativistic heavy-ion collisions (HICs).
In HICs  strongly{\-}-in{\-}teracting matter is created by colliding atomic nuclei at energies much higher than the nuclear mass rest energies. Under such extreme conditions, quarks and gluons are deconfined and form a new phase of quantum chromodynamics (QCD) matter called the quark-gluon plasma (QGP). An extremely important feature of the QGP produced in HICs is that it shows a strong collective behavior and its spacetime evolution can be very accurately described using relativistic hydrodynamics, see, e.g., \cite{Heinz:2013th}. Besides being a nearly perfect relativistic fluid, the QGP exhibits other surprising properties connected to its fluid nature.

Noncentral HICs have large global angular momentum which is estimated to be on the order of thousands of $\hbar$.
It is expected that part of it is transferred to the QGP as vorticity which, in turn, generates particle spin polarization~\cite{Liang:2004ph,Voloshin:2004ha,Betz:2007kg,Becattini:2007sr}. This mechanism resembles the Barnett effect, where a ferromagnet gets magnetized when spinning around an axis~\cite{Barnett:1935}. 
Recent experimental studies showed that some hadrons emitted in noncentral collisions (e.g., Lambda baryons) exhibit a spin alignment along the direction of the global angular momentum. This gives the evidence that the QGP has a strong vortical structure~\cite{STAR:2017ckg,Adam:2018ivw,Acharya:2019vpe}. The global polarization (namely the polarization along the global angular momentum) turns out to be in very good agreement with models proposed in Refs.~\cite{Becattini:2007sr,Becattini:2013vja,Becattini:2013fla,Becattini:2015ska,Becattini:2016gvu,Karpenko:2016jyx,Pang:2016igs,Xie:2017upb}. For recent reviews see, e.g., \cite{Becattini:2020ngo,Liu:2020ymh}.
The assumption of these models is that local thermodynamic equilibrium is reached at some early stage of the process (QGP formation) and kept until hadronization, where the fluid becomes a kinetic hadronic system. At freeze-out, when scatterings cease, particles become polarized only if the thermal vorticity defined in \eq\eqref{thvort} (computed with relativistic hydrodynamics) is different from zero. The formula for the spin vector used to describe the Lambda global polarization is based on an educated ansatz for the distribution function~\cite{Becattini:2013fla}. Such formula is given by the expectation value of
\eq\eqref{plpl3} [or equivalently \eq\eqref{plpl2}] with respect to the local equilibrium state.  After carrying out an expansion in gradients, one obtains at first order~\cite{Becattini:2013fla}
\begin{equation}
\label{spinle}
\langle\hat{\Pi}^\mu(p)\rangle = -\frac{\hbar^2}{8m}\epsilon^{\mu\nu\alpha\beta}p_\nu\frac{\int_{\Sigma_{FO}} d\Sigma_{\lambda}\, p^\lambda f_{F}(1-f_{F}) \varpi_{\alpha\beta}(x)}{\int_{\Sigma_{FO}} d\Sigma_\lambda\, p^\lambda f_{F}},
\end{equation}
where the integration is carried out over the freeze-out hypersurface $\Sigma_{FO}$, $f_F$ is the Fermi-Dirac distribution and $\varpi^{\alpha\beta}$ the thermal vorticity \eqref{thvort}. 

The models which were able to describe so accurately the data~\cite{Becattini:2017gcx,Becattini:2020ngo}, however, fail when it comes to explain the longitudinal Lambda polarization, i.e., the projection of the spin along the beam direction~\cite{Adam:2019srw}. More specifically, the Lambda longitudinal polarization is measured as a function of the azimuthal angle of the transverse momentum and it exhibits a very similar pattern to that of the elliptic flow of the azimuthal particle spectra~\cite{Heinz:2013th}. The predictions of Ref.~\cite{Becattini:2017gcx} for the longitudinal polarization show a correct $\sin( 2\phi)$ behavior, where $\phi$ is the azimuthal angle, but with an opposite sign in the amplitude with respect to the experimental data~\cite{Adam:2019srw}. Unfortunately, this mismatch between theory and experiments, which we will call ``sign puzzle", does not yet have a definitive theoretical explanation, although many attempts have been recently made~\cite{Florkowski:2019qdp,Florkowski:2019voj,Zhang:2019xya,Becattini:2019ntv,Xia:2019fjf,Wu:2019eyi,Sun:2018bjl,Liu:2019krs}. 
A crucial feature of the models in Refs.~\cite{Becattini:2007sr,Becattini:2013vja,Becattini:2013fla,Becattini:2015ska,Becattini:2016gvu,Karpenko:2016jyx,Pang:2016igs,Xie:2017upb} is that they assume local equilibrium also of spin degrees of freedom. 
However, the ``spin puzzle" suggests that spin degrees of freedom may undergo a nontrivial dynamics related to the conversion between orbital and spin angular momentum which is not well understood yet. 

In the past few years, the study of spin dynamics has attracted considerable attention. 
Many works have focused on spin hydrodynamics, an extension of relativistic hydrodynamics where spin degrees of freedom are included. There are several approaches in the literature which, actually, were not specifically developed to address the ``spin puzzle": 
one can promote the total angular momentum conservation as a new hydrodynamic equation of motion with a suitable
 definition of the spin tensor \cite{Florkowski:2017ruc,Florkowski:2017dyn,Florkowski:2018myy,Becattini:2018duy,Florkowski:2018fap,Hattori:2019lfp,Bhadury:2020puc,Weickgenannt:2020aaf}, use the Lagrangian formalism \cite{Montenegro:2017rbu,Montenegro:2017lvf,Montenegro:2018bcf,Montenegro:2020paq} or the holographic duality \cite{Gallegos:2020otk}. 
There has been intense activity also on the description of nonequilibrium dynamics of spin polarization during the collision process using the Wigner-function formalism in the free-streaming case~\cite{Fang:2016vpj,Florkowski:2018ahw,Weickgenannt:2019dks,Gao:2019znl,Hattori:2019ahi,Wang:2019moi,Liu:2020flb,Sheng:2020oqs},
and including particle collisions~\cite{Li:2019qkf,Yang:2020hri,Weickgenannt:2020aaf}.  It is worth to mention that the Wigner-function fomalism has been widely used also for the description of anomalous chiral transport in the QGP, see, e.g.,  Refs.~\cite{Son:2012zy,Hidaka:2016yjf,Hidaka:2017auj,Huang:2018wdl,Gao:2018wmr,Yang:2018lew,Gao:2018jsi,Prokhorov:2018qhq,Carignano:2018gqt,Prokhorov:2018bql,Huang:2020kik}.  
An important question to be addressed is also whether spin equilibrates fast enough for the time scales of HICs.
Calculations of spin equilibration time were  recently carried out using perturbative QCD~\cite{Kapusta:2019sad,Kapusta:2020npk}, Nambu--Jona-Lasinio model~\cite{Kapusta:2019ktm}, and effective vertex for the interaction with thermal vorticity~\cite{Ayala:2019iin,Ayala:2020ndx}.
 In the rest of this section we will discuss the newly developed spin hydrodynamics and quantum kinetic theory as promising approaches for a solution of the ``spin puzzle" and for a deeper understanding of spin effects in HICs. In particular, we will focus on the impact of different pseudo-gauge choices on the formulation of spin hydrodynamics.

\subsection{Spin hydrodynamics and quantum kinetic theory}

\label{secspinhyd}

The equations of motion of conventional relativistic hydrodynamics are the conservation of energy and momentum
\begin{equation}
\label{hydro1}
\partial_\mu T^{\mu\nu}=0 .
\end{equation}
The main idea to extend hydrodynamics to include the dynamics of spin degrees of freedom is by promoting the conservation of the total angular momentum 
\begin{equation}
\label{hydro2}
\partial_\lambda S^{\lambda,\mu\nu} = T^{[\nu\mu]}
\end{equation}
as a new equation of motion, where the spin tensor plays the role of spin density, in the same logic as the energy-momentum tensor is related to energy and momentum density \cite{Florkowski:2017ruc,Florkowski:2017dyn,Florkowski:2018myy,Becattini:2018duy,Florkowski:2018fap,Hattori:2019lfp,Bhadury:2020puc,Weickgenannt:2020aaf}. 
Relativistic hydrodynamics is in principle a classical theory, while spin is inherently a quantum feature of matter. Therefore, the natural starting point for a consistent treatment of spin in hydrodynamics is quantum field theory. In practice, we  establish a connection to quantum field theory by defining our densities $T^{\mu\nu}$ and $S^{\lambda,\mu\nu}$ as ensemble average of quantum operators
\begin{equation}
\label{macro}
T^{\mu\nu} = \langle \hat{T}^{\mu\nu} \rangle , \qquad S^{\lambda,\mu\nu} = \langle \hat{S}^{\lambda,\mu\nu} \rangle .
\end{equation}
The set of equations \eqref{hydro1} and \eqref{hydro2} is called spin hydrodynamics.
The unknowns of this system of equations will be the Lagrange multipliers associated to energy and momentum conservation $\beta^\mu=u^\mu/T$ ($u^\mu$ is the fluid velocity and $T$ the temperature), and to the total angular momentum conservation, the spin potential $\Omega^{\mu\nu}$, introduced in \secc\ref{nonsec}. Moreover, if dissipation effects are considered, extra equations of motion for the dissipative quantities should be provided. In order to compute \eq\eqref{macro}, one has to choose a specific pseudo-gauge.

Relativistic hydrodynamics can be derived, for example, from the Boltzmann equation by applying the method of moments \cite{Denicol:2012cn}. Therefore, having a quantum kinetic theory framework is important to derive spin hydrodynamics. On a microscopic level, angular momentum conservation implies that the conversion between orbital and spin angular momentum can occur only if particles collide with a finite impact parameter. Hence,  we need a kinetic picture where the nonlocality of the collisions is consistently taken into account. In order to formulate a transport theory from quantum field theory, we use the Wigner-function formalism. We define the Wigner function as the normal-ordered ensemble average of the Wigner operator \eqref{wtrans} \cite{DeGroot:1980dk,Heinz:1983nx,Elze:1986qd,Vasak:1987um},
\begin{equation}
W(x,p)\equiv \langle \hat{W}(x,p)\rangle .
\end{equation}
Since we need to introduce collisions, the right-hand side of \eq\eqref{Wignerkin} will be modified as \cite{DeGroot:1980dk}
\begin{equation}
 \left[ \gamma \cdot \left( p+i \,\frac{\hbar}{2} \partial \right) -m\right] W=\hbar\, \mathcal{C} , \label{Wignerkinint}
\end{equation}
where
\begin{equation}
 \C\equiv \int \frac{d^4y}{(2\pi\hbar)^4} e^{-\frac i\hbar p\cdot y}
 \left\langle \mathcal{J}(x_2)\bar{\psi}(x_1)\right\rangle
\end{equation}
and
 $\mathcal{J}= - (1/ \hbar) \partial \mathcal{L}_I /\partial \bar{\psi} $, with $\mathcal{L}_I$ being the interaction Lagrangian. Applying the decomposition \eqref{clifdec} to Eq.~\eqref{Wignerkinint} and separating real and imaginary part, we obtain the equations of motion for the coefficient functions. Thus, we get
 \begin{subequations}
 \label{real2}
\begin{eqnarray}
p\cdot \V -m\F&=& \hbar D_\F ,\label{F2}\\
\frac{\hbar}{2}\partial\cdot  \A+m\Pc&=&-\hbar D_\Pc,\label{P2}\\
p^\mu \F-\frac{\hbar}{2}\partial_\nu \Sc^{\nu\mu}-m\V^\mu&=& \hbar D^\mu_\V ,\label{V2}\\
-\frac{\hbar}{2}\partial^\mu \Pc+\frac12\epsilon^{\mu\nu\alpha\beta}p_\nu S_{\alpha\beta}+m\A^\mu&=&-\hbar D^\mu_\A,\label{A2}\\
\frac{\hbar}{2}\partial^{[\mu} \V^{\nu]}-\epsilon^{\mu\nu\alpha\beta}p_\alpha \A_\beta-m\Sc^{\mu\nu}&=&\hbar D^{\mu\nu}_\Sc ,\label{S2}
\end{eqnarray}
\end{subequations}
for the real part, and 
\begin{subequations}
\label{im2}
\begin{eqnarray}
\hbar\partial \cdot \V&=&2\hbar C_\F ,\label{Vkin2}\\
p \cdot \A&=&\hbar C_\Pc ,\label{orth2}\\
\frac{\hbar}{2}\partial^\mu \F+p_\nu \Sc^{\nu\mu}&=&\hbar C_\V^\mu ,\label{B22}\\
p^{\mu}\Pc+\frac{\hbar}{4}\epsilon^{\mu\nu\alpha\beta}\partial_\nu \Sc_{\alpha\beta}&=& -\hbar C^\mu_\A ,\label{Skin2}\\
p^{[\mu} \V^{\nu]}+\frac{\hbar}{2}\epsilon^{\mu\nu\alpha\beta}\partial_\alpha \A_\beta&=&-\hbar C^{\mu\nu}_\Sc . \label{Akin2}
\end{eqnarray}
\end{subequations}
for the imaginary part, where we defined $D_i = \re \Tr\, (\tilde{\gamma}_i \mathcal{C})$,
$C_i = \im \Tr\, (\tilde{\gamma}_i \mathcal{C})$, $i = \F,\Pc,\V,\A,\Sc$, $\tilde{\gamma}_\F=1$,
$\tilde{\gamma}_\Pc =-i \gamma_5$, $\tilde{\gamma}_\V = \gamma^\mu$,
$\tilde{\gamma}_\A = \gamma^\mu \gamma^5$,
$\tilde{\gamma}_\Sc= \sigma^{\mu \nu}$.
Following Ref.~\cite{Weickgenannt:2020aaf}, we employ
an expansion in powers of $\hbar$ for the  coefficient functions of the Wigner function and
the collision term in Eqs.~\eqref{real2} and \eqref{im2}, e.g., for the scalar part we write
\begin{equation}
	\F = \F^{(0)}+\hbar \F^{(1)} + \mathcal O (\hbar^2 \partial^2).
	\end{equation}
We stress that, since in the equations of motion \eqref{Wignerkinint} a gradient is always accompanied by a factor of $\hbar$, this is effectively also a gradient expansion. We now make the assumption that spin effects are at least of first order in the $\hbar$-gradient expansion. As a consequence it can be shown that~\cite{Weickgenannt:2020aaf} 
\begin{equation}
 \V^\mu=\frac{1}{m}p^\mu \bar{\F}+\mathcal{O}(\hbar^2 \partial^2), \label{newV}
\end{equation}
where we defined 
\begin{equation}
\bar{\F}\equiv\F-\frac{\hbar}{m^2} p^\mu { D_{\V , \mu}^{(1)}}.
\end{equation}
The relevant Boltzmann equations then read
\begin{equation}
 	p\cdot\partial\bar{\F}=m\, C_F ,  \quad
  	p\cdot\partial\A^\mu=m\, C_A^\mu  , \label{Fspintransporttt}
\end{equation}
with $C_F \equiv2C_\F$ and $C_A^\mu \equiv -(1/m) \epsilon^{\mu\nu\alpha\beta}p_\nu C_{\Sc ,  \alpha\beta}$.
In order to obtain a more intuitive understanding of spin-related quantities, we introduce spin as an additional variable in phase space
\cite{Zamanian:2010zz,Ekman:2017kxi,Ekman:2019vrv,Florkowski:2018fap,Bhadury:2020puc,Weickgenannt:2020aaf}. We define the distribution function
 \beq
 \f(x,p,\ms)\equiv\frac12\left[\bar{\F}(x,p)-\ms\cdot\A(x,p)\right],  \label{spinproj}
 \eeq
and the integration measure
%We now employ the quasi-particle approximation, i.e.,
%we assume that $\f$ is of the form
%\begin{equation}
% \f(x,p,\ms)=m\delta(p^2-M^2) f(x,p,\ms)\;, \label{onshellsolution}
%\end{equation}
%where $f(x,p,\ms)$ is a function without singularity at $p^2=M^2 \equiv m^2+ \hbar\delta m^2$ and
%$\delta m^2(x,p,\ms)$ is an interaction contribution to the
%mass-shell condition for free particles
%[where the $\ms$ dependence enters at $\mathcal{O}(\hbar)$]. We introduce the 
%integration measure
\begin{equation}
 \int dS(p)  \equiv \frac{1}{\kappa(p)} \int d^4\ms\, \delta(\ms\cdot\ms+3)\delta(p\cdot \ms),
\end{equation}
with $\kappa(p) \equiv \sqrt{3}\pi/\sqrt{p^2}$
such that
\begin{subequations}
\label{A_int}
\begin{align}
 \bar{\F}&=\int dS(p)\, \f(x,p,\ms), \\
 \A^\mu&= \int dS(p)\, \ms^\mu \f(x,p,\ms).\label{A_int2}
\end{align}
\end{subequations}
The distribution \eqref{spinproj} can be parameterized as
\begin{equation}
 \f(x,p,\ms)=m\delta(p^2-m^2-\hbar\delta m^2) f(x,p,\ms), \label{onshellsolution}
\end{equation}
where $f(x,p,\ms)$ is a function without singularity at $p^2=m^2+ \hbar\delta m^2$ and
$\delta m^2(x,p,\ms)$ is a correction to the
mass-shell condition for free particles arising from interactions.
The final Boltzmann equation to be solved is thus  given by
\begin{equation}\label{Boltzmannnn}
 p\cdot\partial \, \f =m\, \mathfrak{C}[\f] ,
\end{equation}
where $\mathfrak{C}[\f]\equiv \frac{1}{2}(C_F-\ms \cdot C_A)$. 
The collision term $\mathfrak{C}[\f]$ contains both local and nonlocal contributions and has been recently explicitly calculated in Ref.~\cite{Weickgenannt:2020aaf}. It was demonstrated that, using the standard form of the equilibrium distribution function
~\cite{Becattini:2013fla,Florkowski:2017ruc,Florkowski:2018fap}
\begin{equation}\label{f_eq}
 f_{eq}(x,p,\ms)=\frac{1}{(2\pi\hbar)^3}\exp\left[-\beta (x)\cdot p
 +\frac\hbar4 \Omega_{\mu\nu}(x)\Sigma_\ms^{\mu\nu}\right],
\end{equation}
and the total angular momentum conservation in binary scatterings, the conditions under which the collision term vanishes are indeed those of global equilibrium discussed in \secc\ref{nonsec}. In \eq\eqref{f_eq} we defined the spin-dipole-moment tensor $\Sigma_{\ms}^{\mu\nu}\equiv -(1/m) \epsilon^{\mu\nu\alpha\beta}p_\alpha \ms_\beta$.

Once the quantum kinetic theory is established, one can evaluate the hydrodynamic quantities \eqref{macro}. To do so, a pseudo-gauge choice has to be made. Let us first consider the canonical currents. Substituting \eq\eqref{newV} into \eq\eqref{tcanw} and \eq\eqref{A_int2} into \eq\eqref{scanw}, we obtain \footnote{The term $-g^{\mu\nu}(\mathcal{L}_D+\mathcal{L}_I)$ 
in the energy-momentum tensor does not in general vanish when using the equations of motion, but it is proportional to an interaction term. However, we study kinetic theory of dilute systems where such contribution
 to the energy-momentum tensor can be neglected  \cite{DeGroot:1980dk}.}
\begin{eqnarray} \label{canonicalcurrents33}
T_{C}^{\mu\nu}&=& \int dP\, dS\,  p^\mu p^\nu f(x,p,\ms) + \mathcal{O}(\hbar^2\partial^2),\\
 S_{C}^{\lambda,\mu\nu}&=& -\hbar\frac{m}{2}\, \epsilon^{\lambda\mu\nu\alpha}\int dP\, dS\, \ms_\alpha f(x,p,\ms)\n\\
 &=&\hbar \frac{m^2}{2} \int dP\, dS\, \frac{1}{p^2}\left(p^\lambda \Sigma_{\ms}^{\mu\nu}+p^\mu\Sigma_{\ms}^{\nu\lambda}+p^\nu\Sigma_{\ms}^{\lambda\mu}\right)\n\\
&&\times f(x,p,\ms), \label{scanwf}
\end{eqnarray}
where $dP=d^4p \delta(p^2-m^2)$. Note that \eq\eqref{scanwf} is indeed exact.
\footnote{
In Ref.~\cite{Weickgenannt:2020aaf} the spin tensor is defined as the spin tensor in this paper divided by $\hbar$ such that the total angular momentum reads $J^{\lambda, \mu\nu} = x^\mu T^{\lambda\nu}-x^\nu T^{\lambda\mu} + \hbar S^{\lambda,\mu\nu}$. This implies that the $\hbar$ factor in \eq\eqref{scanwf} should not be counted in the $\hbar$-gradient expansion since it is not accompanied by a gradient.
}
Using the Boltzmann equation~\eqref{Boltzmannnn}, the hydrodynamic equations of motion corresponding to the tensors in \eqs\eqref{canonicalcurrents33} and \eqref{scanwf} are given by
\begin{align} \label{Tacan}
 \partial_\mu T^{\mu\nu}_{C} &  = \int dP d S(p)\, p^\nu\,  {\mC}[f]  = 0,\\
\partial_\lambda S_{C}^{\lambda,\mu\nu}
 & = \int dP d S(p)\, \frac\hbar2 \left(\Sigma_\ms^{\mu\nu}\, {\mC}[f]+p^{[\mu}\Sigma_\ms^{\nu]\lambda}\partial_\lambda f(x,p,\ms)\right)\n\\
& = T_{C}^{[\nu\mu]}, \label{total_conscan}
\end{align}
respectively.
Equation \eqref{Tacan} relates the conservation of energy and momentum to the collisional invariant $p^\mu$. On the other hand, from Eq.~\eqref{total_conscan}, which can be viewed as the definition of the antisymmetric part of the energy-momentum tensor, the relation of the divergence of the spin tensor to a collisional invariant is not apparent.
 Furthermore, in global equilibrium, after expanding \eq\eqref{f_eq} up to $\mathcal{O}(\hbar\partial)$ and recalling that $\mathfrak{C}[\f] =0$, \eq\eqref{total_conscan} becomes 
 \begin{align}
 \label{scaneq}
T^{[\mu\nu]}_{C,eq}=\frac{1}{(2\pi\hbar)^3} \frac{\hbar^2}{2} \int dP\, p^{[\nu}\varpi^{\mu]\lambda} p^\rho \varpi_{\lambda\rho} e^{-\beta\cdot p}+\mathcal{O}(\hbar^3\partial^3).
\end{align}
We see from \eqs\eqref{total_conscan} and \eqref{scaneq} that the antisymmetric part of the canonical energy-momentum tensor is different from zero, and hence the spin tensor is not conserved, even in the case of vanishing collisions or global equilibrium. Since the physical picture is that spin changes only due to particle scatterings until global equilibrium is reached, the interpretation of $S_{C}^{\lambda,\mu\nu}$ as a spin density is not consistent.

In Ref.~\cite{Weickgenannt:2020aaf}, it was shown that the HW choice carries interesting physical implications. Starting from the canonical currents and performing the pseudo-gauge transformations in \eqs\eqref{pghw1} and \eqref{pghw2}, one obtains up to first order in the $\hbar$-gradient expansion
\begin{align} \label{KleinGordontensorsss}
 T^{\mu\nu}_{{HW}}={}&\int dP\, dS(p)\,  p^\mu p^\nu f(x,p,\ms) + \mathcal{O}(\hbar^2\partial^2) ,\\
 S^{\lambda,\mu\nu}_{{HW}}
  ={}& \hbar  \int dP\, dS(p) p^\lambda\left(\frac{1}{2} \Sigma_\ms^{\mu\nu}-\frac{\hbar}{4m^2}
  p^{[\mu}\partial^{\nu]}\right) f(x,p,\ms) \n\\
 {}&+ \mathcal{O}(\hbar^2\partial^2). \label{SpinHWw}
 \end{align}
 Notice from \eqs\eqref{canonicalcurrents33} and \eqref{KleinGordontensorsss} that, under our assumption of spin as a first order quantity, the canonical and HW energy-momentum tensor at $\mathcal{O}(\hbar\partial)$ are equal.
The HW hydrodynamic equations of motion can be written with the help of the
Boltzmann equation \eqref{Boltzmannnn} as
\begin{align} \label{Ta}
 \partial_\mu T^{\mu\nu}_{{HW}} &  = \int dP d S(p)\, p^\nu\,  {\mC}[f]  = 0,\\
   \partial_\lambda S_{{HW}}^{\lambda,\mu\nu}
 & = \int dP d S(p)\, \frac\hbar2 \Sigma_\ms^{\mu\nu}\, {\mC}[f]
 = T_{{HW}}^{[\nu\mu]} . \label{total_cons}
\end{align}
Equation \eqref{total_cons} shows the relation between the antisymmetric part of the HW energy-momentum tensor and the collision kernel of the Boltzmann equation. When only local collisions are considered then $\Sigma_{\ms}^{\mu\nu}$ is a collisional invariant, leading to a conserved spin tensor and a vanishing antisymmetric part of the HW energy-momentum tensor in \eq\eqref{total_cons}. In general, when we take into account the nonlocality of the collisions, the spin tensor is not conserved and orbital angular momentum can be converted into spin through the antisymmetric part of the HW energy-momentum tensor which arises at $\mathcal{O}(\hbar^2\partial^2)$. In global equilibrium, the HW energy-momentum tensor is again symmetric and the spin tensor is conserved. 
Therefore, the HW pseudo-gauge turns out to be a consistent choice to describe the conversion between orbital and spin angular momentum of a relativistic fluid. Finally, since the HW and KG spin tensors are the same and, moreover, they differ from the GLW spin tensor by a divergenceless term [see \eqs\eqref{spinwigner} and \eqref{spinglwtrue}], the antisymmetric parts of the HW, KG and GLW energy-momentum tensors are also equal. The differences between these three pseudo-gauges arise at $\mathcal{O}(\hbar^2\partial^2)$, as can be seen from \eqs\eqref{thww}, \eqref{tglww}, \eqref{spinwigner} and \eqref{spinglwtrue} (note that  $\Pc^{(0)}=0$ \cite{Weickgenannt:2020aaf}). The physical implications of these differences require further investigation. {As a final remark for this section, we mention that in Ref. \cite{Weickgenannt:2020aaf} it was shown that, in the nonrelativistic limit, the equations of motion with the HW pseudo-gauge reduce to the well-known form of hydrodynamics with internal degrees of freedom \cite{Lukaszewicz1999}. 
}

\section{Pseudo-gauge transformations and the relativistic center of inertia}
\label{centsec}

In Newtonian mechanics the center of mass has an unambiguous definition: it is the unique point obtained by the mean of all points weighted by the local mass. However this is not true anymore for a relativistic system, since the energy (or inertial mass) depends on the velocity. Following Ref.~\cite{Pryce:1948pf}, we discuss several possible definitions for the relativistic center of mass and study their physical meanings when internal angular-momentum degrees of freedom are included (see also a recent related work Ref. \cite{Lorce:2018zpf}). {In particular, we show that, since a pseudo-gauge transformation can be seen as a rearrangement of the splitting between spin and orbital angular momentum, the different choices of spin tensors discussed in the previous sections can be related to different definitions of the relativistic center of mass. }

\subsection{External and internal components of angular momentum}

Let us consider for simplicity classical fields.
In special relativity, any angular momentum can be decomposed into an internal and external part with respect to a reference point $X^\mu$,
\begin{equation}
J^{\mu\nu}=L_X^{\mu\nu}+ S_X^{\mu\nu}, \label{exinam}
\end{equation}
where the external component is given by
\begin{equation}
\label{lxx}
L^{\mu\nu}_X\equiv X^{[\mu} P^{\nu]}
\end{equation} 
and the internal component $S_X^{\mu\nu}$ describes the rotation about $X^\mu$. The term $S_X^{\mu\nu}$ is not necessarily related to spin in the corresponding quantum theory. On the other hand, we can also decompose the total angular momentum into generators of boosts $K_n^\mu$ and generators of rotation $J_n^\mu$ depending on the four-velocity $n^\mu$ of the frame in which the generators are defined,
\begin{equation}
J^{\mu\nu}= K_n^{[\mu} n^{\nu]}-\epsilon^{\mu\nu\alpha\beta}n_\alpha J_{n\beta} , \label{boorotam}
\end{equation}
with $K_n^\mu\equiv J^{\mu\nu} n_\nu$ and $J_n^\mu\equiv -\frac12 \epsilon^{\mu\nu\alpha\beta}n_\nu J_{\alpha\beta}$. Combining Eqs.~\eqref{exinam} and \eqref{boorotam} we obtain
\begin{equation}
J^{\mu\nu}= (K_{n,L_X}^{[\mu}+K_{n,S_X}^{[\mu}) n^{\nu]}-\epsilon^{\mu\nu\alpha\beta}n_\alpha (J_{n,L_X , \beta}+J_{n,L_X , \beta}) , \label{exinboorotam}
\end{equation}
where
\begin{subequations}
\label{subde}
\begin{eqnarray}
K_{n,L_X}^{\mu}&\equiv&(P\cdot n) X^\mu- (X\cdot n) P^\mu  , \label{de11}\\
K_{n,S_X}^{\mu}&\equiv& S_X^{\mu\nu} n_\nu, \label{de22} \\
J_{n,L_X}^{\mu}&\equiv& -\epsilon^{\mu\nu\alpha\beta}X_\nu P_\alpha n_\beta  ,\\
J_{n,S_X}^{\mu}&\equiv& -\frac12\epsilon^{\mu\nu\alpha\beta} n_\nu S_{X , \alpha\beta}. \label{dedede}
\end{eqnarray}
\end{subequations}
Consequently, if we want to identify the internal angular momentum with the generators of rotation in the frame characterized by the four-velocity $n^\mu$, we should impose the condition
\begin{equation}
S_X^{\mu\nu} n_\nu=0 \label{frenkel_intrinsic}
\end{equation}
in order to remove the contribution from the boost generators to the internal part in \eqs\eqref{subde}. In this case, \eq\eqref{dedede} can be inverted to obtain the internal angular momentum in terms of the rotation generators
\begin{equation}
S_X^{\mu\nu}=-\epsilon^{\mu\nu\alpha\beta}n_\alpha J_{n,S_X , \beta}.
\end{equation}

\subsection{Center of inertia and centroids}

A natural definition of the center of inertia of a system is the mean of all points weighted by the local energy. Given an energy-momentum tensor $T^{\mu\nu}$ in a certain pseudo-gauge we define the center of inertia as
\begin{equation}
 q^\mu\equiv \frac{1}{P^0} \int d^3x\, x^\mu T^{00}=\frac{1}{P^0}(x^0 P^\mu+ L^{\mu0}) , \label{inertia}
\end{equation}
where $L^{\mu\nu}\equiv J^{\mu\nu}-S^{\mu\nu}$ and $P^\mu = \int d^3x T^{0\mu}$. The definition above implies $q^0=x^0$. Using the conservation of the energy-momentum tensor, we obtain for the time derivative of the center of inertia
\begin{equation}
\label{th1}
\partial_{0}q^\mu=\frac{1}{P^0}\int d^3x\, T^{\mu0},
\end{equation}
provided that, as usual, boundary terms can be neglected.
If we require that the center of inertia moves along a straight line, then we need to impose
\begin{equation}
\label{condmass}
\partial_\nu T^{\mu\nu}=0,
\end{equation}
since in this case 
\begin{equation}
(\partial_{0})^2 q^\mu=\frac{1}{P^0}\int d^3x\, \partial_\nu T^{\mu\nu}=0 . \label{comt}
\end{equation}
We call \eq\eqref{th1} together with \eq\eqref{condmass} the relativistic center-of-mass theorem.
The condition \eqref{condmass} is trivially fulfilled for symmetric energy-momentum tensors, e.g.,  for the Belinfante, HW, GLW and KG pseudo-gauge. We note from \eq\eqref{nccan} that the condition~\eqref{condmass} is also fulfilled for the canonical case, since the canonical spin tensor is totally antisymmetric. Hence, all energy-momentum tensors discussed in this paper are consistent with the relativistic center-of-mass theorem.

%We can understand pseudo-gauge transformation as a redefinition of the center of mass in the following way \cite{Pryce:1948pf}: We start from the total angular momentum
%\begin{equation}
% M^{\mu\nu}\equiv \int d^3x\, \left(x^\mu T^{0\nu}_{Bel}-x^\nu T^{0\mu}_{Bel}\right).
%\end{equation}
%We define the canonical mean position $q^\mu$ in the lab frame as
%\begin{equation}
% P^0 q^\mu\equiv \int d^3x\, x^\mu T^{00}=x^0 P^\mu+ M^{\mu0}.
%\end{equation}

Obviously the center of inertia \eqref{inertia} is not covariant. It can be generalized in a covariant way by introducing the so-called centroid $q_n^\mu$ which is identical to the center of inertia in a generic frame moving with four velocity $n^\mu$,
\begin{equation}
 q_n^\mu=\frac{1}{P\cdot n} (x^0_n P^\mu + L^{\mu\nu}n_\nu), \label{centroid}
\end{equation}
where $x_n^0\equiv x\cdot n$ is the time in the given frame. Notice also that  $q_n \cdot n = x^0_n$.

Now we want to express the orbital angular momentum as
\begin{equation}
\label{lqq}
L^{\mu\nu}=q_n^{[\mu}P^{\nu]},
\end{equation}
which implies the choice of the centroid as reference point, $X^\mu\equiv q_n^\mu$, and hence, from \eq\eqref{lxx}, $L^{\mu\nu}=L^{\mu\nu}_{q_n}$.
 As a consequence, from \eq\eqref{exinam}, we identify the global spin with the internal angular momentum $S^{\mu\nu} = S^{\mu\nu}_{q_n}$.
We also note that the condition \eqref{frenkel_intrinsic} for the spin tensor is needed to ensure the validity of the center-of-mass theorem \eqref{comt} and to make the centroid a Lorentz vector, since it allows us to write it in terms of conserved quantities
\begin{equation}
 q_n^\mu=\frac{1}{P\cdot n} (x^0_n P^\mu + J^{\mu\nu}n_\nu). \label{centroidframen}
\end{equation}
Requiring $q_n^0 = q^0= x^0$, we obtain $x^0_n= x^0\frac{P\cdot n}{P^0}-\frac{1}{P^0}L^{0\nu}n_\nu$ and thus
\begin{equation}
 q_n^\mu=\frac{x^0 P^\mu}{P^0}-\frac{L^{\nu0}n_\nu P^\mu}{P^0(P\cdot n)}+\frac{L^{\mu\nu}n_\nu}{P\cdot n},
\end{equation}
in order to obtain the worldline of $q_n$ parametrized by the original time coordinate $x^0$ and to compare to \eq\eqref{inertia}.

We stress that writing the orbital part in terms of the centroid as in \eq\eqref{lqq} is not possible for all pseudo-gauges. In the next subsections we will study whether we can establish a correspondence between different choices of $n^\mu$ and different expressions for $S^{\mu\nu}$.

\subsection{Belinfante pseudo-gauge}

Since in the Belinfante case the spin tensor vanishes, we have $L_B^{\mu\nu}=J^{\mu\nu}$. Thus
\begin{equation}
 q^\mu=\frac{1}{P^0}(x^0 P^\mu+ J^{\mu0}) \label{inertia_Bel}
\end{equation}
and
\begin{equation}
 q_n^\mu=\frac{1}{P\cdot n} (x^0_n P^\mu + J^{\mu\nu}n_\nu).
\end{equation}

\subsection{Center of inertia as reference point: Canonical pseudo-gauge} \label{subsec_inertia}

The canonical global spin fulfills the condition \eqref{frenkel_intrinsic} in the frame specified by $n^\mu=(1,{\bf 0})$, i.e., we have $S_C^{i0}=0$ (see \secc\ref{pl_sec}).
We can then use the canonical currents to evaluate Eq.~\eqref{inertia} and obtain
\begin{equation}
 q^\mu=\frac{1}{P^0}(x^0 P^\mu+ L_C^{\mu0})=\frac{1}{P^0}(x^0 P^\mu+ J^{\mu0}). \label{inertia_Can}
\end{equation}
We can now define the global spin
\begin{equation}
 S_{q}^{\mu\nu}\equiv J^{\mu\nu}-q^{[\mu} P^{\nu]} \label{MDspin}
\end{equation}
which, as the canonical spin, fulfills $S_{q}^{i0}=0$ in any frame and is not a tensor. 
We stress that $S_{q}^{\mu\nu}$ is different from $S^{\mu\nu}_C$ as $L^{\mu\nu}_C$ cannot be expressed in terms of the center of inertia $q^\mu$, even though the canonical currents were used to calculate $q^\mu$. Defining the spatial components of the total angular momentum $J^{ij}\equiv \epsilon^{ijk}J^k$ we obtain
\begin{equation}
\label{smd}
\boldsymbol{S}_{q}=\mathbf{J}-\mathbf{q}\times \mathbf{P}.
\end{equation}
If we want to go from a classical to a quantum framework where $P^\mu$ as well as $q^\mu$ are promoted to be operators, then \eq\eqref{smd} is given by \cite{Pryce:1948pf}
\begin{equation}
\hat{\boldsymbol{S}}_{q}=\int d^3x\, \frac{\hbar}{2(\ppm^0)^2}\psi^\dagger\left[m^2 \boldsymbol{\mathfrak{S}}+i m\, \pp \times \boldsymbol{\gamma} +(\pp \cdot\boldsymbol{\mathfrak{S}}) \pp\right]\psi, \label{MD_explicit}
\end{equation} 
provided that the operators act on a single-particle state with momentum $\ppm^\mu$ and with  $\boldsymbol{\mathfrak{S}}$ defined in \eq\eqref{bolds}.

\subsection{Center of mass as reference point: HW, GLW and KG pseudo-gauges}

For a system with finite mass $m$ there is a preferred reference frame for defining physical quantities in a covariant way in terms of the Poincar{\'e} generators. This frame is given by the comoving frame of the system, denoted by the four-velocity $n_\star^\mu\equiv P^\mu/m$ (the subscript $\star$ indicates that the corresponding quantities are evaluated in the rest frame). Clearly, in this frame the mass is given by
\begin{equation}
m\equiv P^0_\star= P\cdot n_\star.
\end{equation}
The corresponding centroid, that we call the center of mass, is obtained by using $n^\mu_\star$ in Eq.~\eqref{centroid}, i.e., 
\begin{equation}
 q_\star^\mu=\frac{1}{m} \left(\tau P^\mu + \frac1m J^{\mu\nu}P_\nu\right),
\end{equation}
where we defined the proper time $\tau\equiv x^0_\star$ and already imposed $P_\mu S^{\mu\nu}_\star=0$ in accordance with \eq\eqref{frenkel_intrinsic}. Note that $q_\star^\mu$ is a Lorentz vector.
In a similar way, the spin of a massive particle is defined as the proper internal angular momentum, i.e., choosing the reference vector $n^\mu_\star$ and the reference point $q_\star^\mu$ in Eqs.~\eqref{subde}. This leads to
\begin{subequations}
\label{subde2}
\begin{eqnarray}
K_{\star,L}^{\mu}&=& m q_\star^\mu- \tau P^\mu,\\
K_{\star,S}^{\mu}&=& 0,\\
J_{\star,L}^{\mu}&=& 0,\\
J_{\star,S}^{\mu}&=& -\frac{1}{2m}\epsilon^{\mu\nu\alpha\beta} P_\nu S_{\star\alpha\beta}. \label{dedede_rf}
\end{eqnarray}
\end{subequations}
In the last equation, we identify the Pauli-Lubanski vector $w^\mu=mJ_{\star,S}^{\mu}$ (cf. with \secc\ref{plvec}), and we verify that it is indeed  identical to the generator of rotations defined in the center-of-mass frame. 

One can show that the HW (hence the GLW and KG) global spin \eqref{stothw} is obtained as the difference between the total angular momentum and the orbital angular momentum with respect to the center of mass, i.e.,
\begin{equation}
S_{HW}^{\mu\nu}=S_{q_\star}^{\mu\nu}\equiv  J^{\mu\nu}-q_\star^{[\mu }P^{\nu ]} ,
\end{equation}
which is clearly a Lorentz tensor. The HW spin vector  is then given by
\begin{equation}
\label{shwcent}
\boldsymbol{S}_{HW}=\mathbf{J}-\mathbf{q}_\star\times \mathbf{P},
\end{equation}
which corresponds to \eq\eqref{shwgf}
and, in the rest frame,
\begin{equation}
\boldsymbol{S}_{HW\star}=\mathbf{J}=\boldsymbol{S}_C,
\end{equation}
consistently with \eq\eqref{hwcc}~\cite{Pryce:1948pf}.

It is worth mentioning that there is another option for the reference point, namely the mean position~\cite{Pryce:1948pf}
  \begin{equation}
   \tilde{q}^\mu\equiv \frac{1}{P^0+m}(P^0 q^\mu+m q_\star^\mu),
  \end{equation}
     which is not a vector.
The internal angular momentum with respect to this point corresponds to the internal angular momentum
\begin{equation}
S_{\tilde{q}}^{\mu\nu}\equiv J^{\mu\nu}- \tilde{q}^{[ \mu} P^{\nu]} ,
\end{equation}
with spin vector
\begin{equation}
 \boldsymbol{S}_{\tilde{q}}=\mathbf{J}-\mathbf{\tilde{q}}\times \mathbf{P}.
\end{equation}
After quantizing by promoting $P^\mu$ and $\tilde{q}^\mu$ to operators~\cite{Pryce:1948pf}, the spin vector reads
\begin{align}
\label{spifw}
\hat{\boldsymbol{S}}_{\tilde{q}}=&\int d^3x\, \frac{\hbar}{2\ppm^0}\psi^\dagger\left[m\, \boldsymbol{\mathfrak{S}}+i\,\pp\times \boldsymbol{\gamma}\right.\n\\
&\left.+\frac{1}{\ppm^0+m}(\pp\cdot\boldsymbol{\mathfrak{S}})\pp\right]\psi,
\end{align}
where $\pp$ is again the three-momentum of the one-particle state. Equation~\eqref{spifw} corresponds to the spin vector derived by Foldy and Wouthuysen in Ref.~\cite{PhysRev.78.29}.

We conclude this subsection with a physical remark.
The canonical currents describe position and spin emerging directly from the Dirac equation and thus contain the rapid oscillation (``Zitterbewegung") in the motion of a Dirac particle. However, this oscillation is not measurable \cite{schrodinger1930kraftefreie,schrodinger1931quantendynamik,Dirac1930} and should be removed to obtain physical quantities. Equations \eqref{MD_explicit}, \eqref{shwgf} and \eqref{spifw} are expressions calculated from various definitions of the relativistic center of mass. As these definitions are mean positions, the spin vectors \eqref{MD_explicit}, \eqref{shwcent} and \eqref{spifw} do not contain the rapid oscillation \cite{Pryce:1948pf}.

\subsection{Massless particles and side jumps}

We summarize the results obtained in this section for the massive case as follows: the splitting of the total angular momentum into orbital and spin part 
\begin{equation}
J^{\mu\nu}= q_n^{[\mu} P^{\nu]}+S^{\mu\nu}
\end{equation}
can be fixed by requiring 
\begin{equation}
\label{cond123}
n_\mu S^{\mu\nu}=0 
\end{equation}
as a supplementary condition, which determines $q_n^\mu$ according to Eq.~\eqref{centroidframen}. For finite mass, choosing $n^\mu=n_\star^\mu$ as a frame vector yields a unique covariant decomposition corresponding to the HW pseudo-gauge. 

For vanishing mass, however, the absence of a rest frame leads to additional complications. As $P^2=0$, imposing the condition \eqref{cond123} with $n^\mu \propto P^\mu$, $P_\mu S^{\mu\nu}=0$, does not determine the splitting uniquely. Consider a redefinition of the position $q_P^\mu \to \tilde{q}_P^\mu = q_P^\mu+\Delta^\mu$ with a shift $\Delta^\mu$, then we need to redefine a new global spin as $S^{\mu\nu}\to \tilde{S}^{\mu\nu} = S^{\mu\nu} - \Delta^{[\mu} P^{\nu]} $ in order for the total angular momentum to be the same. The condition on the new global spin $P_\mu \tilde{S}^{\mu\nu}=0$ holds if $P_\mu \Delta^\mu =0$. A solution to this condition can be found such that $\Delta^\mu$ is not proportional to $P^\mu$ leading to $\Delta^{[\mu} P^{\nu]}\neq 0$. This implies that the definition of orbital and spin angular momentum is ambiguous
\cite{Chen:2014cla,Chen:2015gta,Stone:2014fja,Stone:2015kla}. Thus, in contrast to the massive case, there is no possibility to determine the spin in a frame-independent way. In other words, the HW pseudo-gauge, which is related to the particle rest frame, does not exist. This fact is also apparent from \eqs\eqref{shw} and \eqref{shwgf} as a factor of $m$ is present in the denominator. It may  seem natural to use the  canonical spin tensor instead. However, Eq.~\eqref{canspinv} does not yield a familiar definition for the spin of a massless state since it should be slaved to the particle momentum. Interestingly, we note that the quantum spin vector in \eq\eqref{MD_explicit} has a smooth massless limit which yields the familiar form related to the particle helicity
\begin{equation}
\hat{\boldsymbol{S}}_{q,m=0}=\int d^3x\, \frac{\hbar}{2}\psi^\dagger \frac{\pp\cdot\boldsymbol{\mathfrak{S}}}{|\pp |}\frac{\pp}{|\pp |}\, \psi = \hbar\, \lambda \frac{\pp}{|\pp |}, \label{MD_explicit_massless}
\end{equation} 
where we considered the expectation value of a one-particle state $\ket{\ppm , \lambda}$ with helicity $\lambda=\pm 1/2$.
Hence, we can generalize
\begin{equation}
S^{\mu\nu}_{q,m=0}=\hbar\, \lambda \frac{1}{\ppm \cdot \bar{n}} \epsilon^{\mu\nu\alpha\beta}\ppm_\alpha \bar{n}_\beta, \label{spinqnbar}
\end{equation}
where $\bar{n}^\mu\equiv(1,{\bf 0})$ in any frame. This coincides with the global spin used in Ref.~\cite{Chen:2015gta} and, as pointed out in \secc\ref{subsec_inertia}, corresponds to defining the position as  the center of inertia $q^\mu$, which is not a Lorentz vector. As a consequence, a Lorentz transformation leads to a shift in the position known as the side-jump effect \cite{Chen:2014cla,Chen:2015gta,Stone:2014fja,Duval:2014ppa,Stone:2015kla,Huang:2018aly}. Consider a Lorentz transformation $\Lambda$, the total angular momentum is a tensor and will transform as 
\begin{equation}
J^{\mu\nu}\to J^{\prime\mu\nu}=\Lambda^\mu_\alpha \Lambda^\nu_\beta J^{\alpha\beta}. \label{JLT}
\end{equation}
 On the other hand, the spin \eqref{spinqnbar} transforms as
\begin{align}
S^{\mu\nu}_{q,m=0}\rightarrow S^{\prime\mu\nu}_{q,m=0}&=\hbar\, \lambda \frac{1}{\ppm^\prime\cdot \bar{n}} \epsilon^{\mu\nu\alpha\beta}\ppm^\prime_\alpha \bar{n}_\beta\n\\
&= \hbar\, \lambda \frac{1}{\ppm^\prime \cdot n^\prime} \epsilon^{\mu\nu\alpha\beta}\ppm^\prime_\alpha n^\prime_\beta- \Delta^{[\mu} \ppm^{\prime \nu]}\n\\
&= \Lambda^\mu_\alpha \Lambda^\nu_\beta S^{\alpha\beta}_{q,m=0}- \Delta^{[\mu} \ppm^{\prime \nu]} \label{spinltan}
\end{align}
with $\ppm^{\prime\mu}\equiv \Lambda^\mu_\nu \ppm^\nu$ and $n^{\prime\mu}\equiv \Lambda^\mu_\nu \bar{n}^\nu$. Here the term $\Delta^{[\mu} \ppm^{\prime \nu]}$ is an anomalous contribution to the Lorentz transformation of the global spin for Eq.~\eqref{spinqnbar}  to be preserved after the Lorentz transformation. In order to ensure Eq.~\eqref{JLT}, 
\begin{align}
J^{\prime\mu\nu}&= q^{\prime[\mu} \ppm^{\prime \nu]}+S^{\prime\mu\nu}_{q,m=0}\n\\
&= \Lambda^\mu_\alpha \Lambda^\nu_\beta J^{\alpha\beta}
\end{align}
the center of inertia $q^\mu$ has to transform as
\begin{align}
q^\mu\to q^{\prime\mu}= \Lambda^\mu_\nu q^\nu+\Delta^\mu.
\end{align}
The anomalous contribution of a Lorentz transformation for the center of inertia $\Delta^\mu$ can be found by contracting Eq.~\eqref{spinltan} with $\bar{n}_\nu$. Choosing $\Delta^\mu$ to be purely spatial in the frame at rest with the observer after the Lorentz transformation, i.e., $\bar{n}\cdot \Delta=0$, we obtain in this frame
\begin{equation}
\label{sshift}
\Delta^{\mu}=\hbar\lambda\frac{\epsilon^{\mu\nu\alpha\beta}\ppm^\prime_\nu n^\prime_\alpha\bar{n}_\beta}{(\ppm^\prime\cdot \bar{n})(\ppm^\prime\cdot n^\prime)}.
\end{equation}
The physical implications of the anomalous shift $\Delta^{\mu}$ can be seen in a binary particle scattering $\ppm_{1i}+\ppm_{2i} \to \ppm_{1f}+\ppm_{2f}$. Consider first the frame, called ``no-jump frame", which we assume to coincide with the center-of-momentum frame, where the two initial particles collide in one point and the final particles are emitted from the same point. If we see the scattering in a boosted frame in a direction parallel to the initial momenta, then we have to compute the shift $\Delta^\mu$ in \eq\eqref{sshift} for each particle. For the two incoming particles, since the spatial components of $n^{\prime\mu}$ are parallel to the three-momenta, then $\Delta^\mu_{1i}=\Delta^\mu_{2i}=0$. For the final particles, since the momenta are not parallel to the spatial components of $n^{\prime\mu}$ anymore, we have that  $\Delta^\mu_{1f}$ and $\Delta^\mu_{2f}$ are different from zero. This means that the particles in the final state are emitted in a position shifted by an amount $\Delta^\mu_{1f}$ and $\Delta^\mu_{2f}$, respectively, from the point where the initial particles collided. This is the side-jump effect.

We stress that the side jump effect occurs for massless particles due to the absence of a covariant definition of the center of mass and, hence, of a covariant spin. For massive particles, instead, it is always possible to define a covariant center of mass which leads to the HW, GLW or KG spin {(at least for free fields or in case of local interactions). Therefore, in the massive case it is natural to use the HW, GLW or KG pseudo-gauge where the spin is defined in the particle rest frame and no anomalous shift has to be taken into account.
In the case of nonlocal interactions, the HW spin tensor turns out not to be conserved anymore [see Eq. \eqref{total_cons}] and the situation might be different. However, for a full understanding of this issue, further studies are needed.}

\section{Einstein-Cartan theory}
\label{grsec}

{So far, we have studied different energy-momentum and spin tensors in flat spacetime. However, as in gravitational physics the energy-momentum tensor is directly measurable through geometry, it is also interesting to review the role of the densities in curved spacetime.}
In conventional general relativity the energy-momentum tensor is defined 
 following Belinfante and Rosenfeld as~\cite{belinfante1939spin,belinfante1940current,rosenfeld1940energy}
\begin{equation}
\label{tgr}
T_{\mu\nu}=\frac1 g \, \frac{\delta {A}_M}{\delta g^{\mu\nu}},
\end{equation}
where the matter action $A_M=\int d^4x \mathcal{L}_M$, with $\mathcal{L}_M$ the Lagrangian, $g^{\mu\nu}$ is the metric tensor and $g=\sqrt{-\text{det}(g^{\mu\nu})}$. In the literature, the expression for the energy-momentum tensor above is often considered to be the fundamental one because it is defined as the source of the gravitational field. It is important to note that since $g^{\mu\nu}$ is a symmetric tensor, $T_{\mu\nu}$ in \eq\eqref{tgr} is also symmetric and indeed reduces in special relativity to the energy-momentum tensor discussed in \secc\ref{Bel:sec}. Following these considerations, it is usually claimed that the ``physical" energy-momentum tensor must be symmetric. However, we observe that in conventional general relativity spinorial degrees of freedom are not taken into account and we can regard the absence of an antisymmetric part of the energy-momentum tensor as the consequence of this fact. In the following we shall briefly review an extension of general relativity called Einstein-Cartan theory, where one allows the spacetime geometry to have a nonvanishing torsion, an additional property of the manifold geometry which couples to spin. In such theory the energy-momentum tensor can gain an antisymmetric part~\cite{hehl1976general,hehl1980four,blagojevic2001gravitation}.

\subsection{Riemann-Cartan geometry}

The Einstein-Cartan theory is based on the so-called Riemann-Cartan spacetime~\cite{hehl1976general,hehl1980four,Hehl:1984ev,blagojevic2001gravitation,Hehl:2007bn}. Let us consider a four-dimensional differentiable manifold, whose spacetime points are labeled with $x^\mu$. In order to specify the geometrical structure, we introduce a spacetime dependent symmetric metric $g^{\mu\nu}=g^{\mu\nu}(x)$ (such that $g^{\mu\alpha}g_{\alpha\nu}=\delta^\mu_\nu$, with $\delta^\mu_\nu$ the Kronecker delta) and the notion of parallel transport of vectors. Consider an infinitesimal displacement $x^\mu+dx^\mu$ from the point $x^\mu$, then a vector $B^\mu$ changes by
\begin{equation}
d B^\mu = -\tilde{\Gamma}^{\mu}_{\alpha\beta}(x) B^\alpha dx^\beta ,
\end{equation}
where $\tilde{\Gamma}^{\alpha}_{\beta\mu}$ is the affine connection. In contrast to conventional Einstein's general relativity, we allow the affine connection to have an antisymmetric part of the form
\begin{equation}
 F_{\alpha\beta}^{\ \ \ \mu\ }\equiv \frac12\tilde{\Gamma}_{[\alpha\beta]}^{\mu},
\end{equation}
which is called torsion tensor. If we constrain the affine connection in such a way that the covariant derivative vanishes (metric compatibility), i.e., impose local Minkowski structure, we can write the affine connection as
\begin{equation} \label{affinecon}
 \tilde{\Gamma}_{\alpha\beta}^{\mu}=\Gamma_{\alpha\beta}^{\mu}-K_{\alpha\beta}^{\ \ \ \mu} ,
\end{equation}
where 
\[
\Gamma_{\alpha\beta}^{\mu} = \frac 12 g^{\mu\nu}(\partial_\alpha g_{\beta\nu}+\partial_\beta g_{\alpha\nu}-\partial_\nu g_{\alpha\beta})
\]
is the conventional Christoffel symbol and
\begin{equation}
\label{cont}
 K_{\alpha\beta}^{\ \ \ \mu}=-F_{\alpha\beta}^{\ \ \ \mu}+F_{\beta\ \,\alpha}^{\ \,\mu}-F^\mu_{\ \,\alpha\beta}
\end{equation}
is the contorsion tensor. Obviously, if torsion vanishes we recover the usual Riemann spacetime. While curvature can be regarded as a ``rotation field strength" connected to the loss of parallelism of parallel transported
vectors, torsion can be interpreted as a ``translation field strength" which is manifest in the closure failure of parallelograms~\cite{Hehl:2007bn}.

\subsection{Tetrads and spinors in curved spacetime}

Given the metric $g_{\mu\nu}$ defined on the manifold, we can always define a tangent space at each spacetime point and establish a local flat orthonormal coordinate system $e_a(x)=e_{\ \, a}^{\mu}(x)\partial_\mu$ which are called tetrads or vierbein.~\footnote{We use the Greek letters to denote the conventional holonomic spacetime indices and the Latin letters $a,b,...=0,1,2,3$ for the anholonomic tangent-space indices.}  Hence, their components $e_{\ \, a}^{\mu}$ and the reciprocal $e^{\ \, a}_{\mu}$ are such that
\begin{equation}
e_{\ \, a}^{\mu} \,e^{\ \, a}_{\nu}=\delta^\mu_\nu, \qquad e_{\ \, a}^{\mu} \,e^{\ \, b}_{\mu}=\delta^b_a
\end{equation}
and 
\begin{equation}
g_{\mu\nu}=e^{\ \, a}_{\mu} \, e^{\ \, b}_{\nu} \,g_{ab} ,
\end{equation}
where $g_{ab}=(+1,-1,-1,-1)$ is the Minkowski metric. Spacetime indices are raised or lowered with $g^{\mu\nu}$, tetrad indices with $g_{ab}$, and transvection is done by appropriate contraction with the tetrads, e.g., for a vector $B_\mu$, we define $B_a=e_{\ \, a}^{\mu}B_\mu$.
The intuitive picture is that we are assigning at each spacetime point an observer which measures lengths and time with respect to the local flat coordinate system $e_a$.

We now introduce a classical spinor $\psi$ which will play the role of matter field. In the locally flat spacetime all the familiar properties of the spinors hold, {in particular they transform under a Lorentz transformation of the tetrads $e^\mu_{\ \, a}\rightarrow \Lambda_a^{\  b} e^\mu_{\ \, b}$ as $\psi\rightarrow U(\Lambda)\psi$ with $U^{-1}\gamma^a U=\Lambda^a_{\ b}\gamma^b$ and all the conventional relations of the Dirac $\gamma$-matrices apply.} The covariant derivative of a spinor is defined as
\begin{equation}\label{covder}
 D_\mu \ps\equiv \left(\partial_\mu -\frac12 \omega_\mu^{\ \, ab}f_{ab}\right	) \ps ,
\end{equation}
where $f_{ab}=-\frac i 2 \sigma_{ab}$ and $\sigma_{ab}$ is given by \eq\eqref{sig}. The quantity $\omega_\mu^{\ \, ab}=-\omega_\mu^{\ \, ba}$ is the spin connection
\begin{equation}\label{spinconnection}
 \omega_{\mu a}^{\ \ \,b}=\frac12(-\Omega_{\mu a}^{\ \ \, b}+\Omega_{a\ \mu}^{\ b}-\Omega^b_{\ \mu a}-K_{\mu\nu}^{\ \ \,\alpha}e_\alpha^{\ \, b} e^\nu_{\ \, a}) ,
\end{equation}
with $
 \Omega_{\mu\nu}^{\ \ \, a}\equiv \partial_{[\mu} e_{\nu]}^{\ \, a}$ and $K_{\mu\nu}^{\ \ \, \alpha}$ being the contorsion in \eq\eqref{cont}. 
%The Dirac equation reads 
%\begin{equation}
%(i \gamma^a e^{\mu}_{\ \, a}D_\mu - m)\psi=0
%\end{equation} 
The commutator of the covariant derivatives is given by
 \begin{equation}
 [ D_\mu , D_\nu ]\psi = -\frac12 R_{\mu\nu}^{\ \ \,\,ab}f_{ab}\psi,
 \end{equation}
where $R_{\mu\nu}^{\ \ \,\,ab}$ is the Riemann-Cartan tensor
\begin{equation}
R_{\mu\nu}^{\ \ \,\,ab}=\partial_\mu \omega_{\nu}^{\ \, ab}-\partial_\nu \omega_{\mu}^{\ \, ab} + \omega_{\mu c}^{\ \, \ 	\, b}\omega_{\nu}^{\ \, a c} - \omega_{\nu c}^{\ \, \ 	\, b}\omega_{\mu}^{\ \, a c} .
\end{equation}
The inclusion of torsion will also lead to a modification of the field equations~\cite{hehl1976general}.

\subsection{Local Poincar{\'e} transformations and conservation laws}

The fundamental idea of the Einstein-Cartan theory is to promote the global Poincar{\'e} symmetry of the action to a gauge symmetry~\cite{hehl1976general,hehl1980four,blagojevic2001gravitation}. This approach is analogous to the Yang-Mills formulation of gauge theory. 

In order to give as an intuitive explanation as possible of the Einstein-Cartan theory, we will start by considering a flat Minkowski spacetime. In this case, the tetrads will simply be
\begin{equation}
e^\mu_{\ \, a}=\delta^\mu_a
\end{equation}
and what is discussed in \secc\ref{secvarious} holds, expect here we restrict to classical fields. The global Poincar{\'e} transformations \eqref{stt} and \eqref{ptlor}, which we write in a compact form 
\begin{equation}
x^\mu \to x^{\prime \mu} = x^\mu + \zeta^{\mu\nu}x_\nu + \xi^\mu ,
\end{equation}
will induce a functional variation of the spinor  
\begin{equation}
\label{functvar}
\delta\psi = \psi^\prime (x)-\psi (x) = \left(\frac12\zeta^{	ab}f_{ab}-\Xi^a \partial_a\right)\psi (x) ,
\end{equation}
with $\Xi^a=\xi^a+\zeta_{\ \,\, b}^{a}\delta^b_\mu x^\mu$.
Using \eq\eqref{functvar} we obtain through  Noether's theorem the canonical currents. Let us now promote the ($4+6$) infinitesimal parameters of the Poincar{\'e} transformations to be functions of spacetime, $\xi^a (x)$ and $\zeta^{ab}(x)$. If we now calculate the variation of the action \eqref{action}, $\delta A$, with respect to these new local Poincar{\'e} transformations and make use of the spinor variation in \eq\eqref{functvar}, we obtain~\cite{hehl1976general}
\begin{align}
\delta A&= \int d^4x \left[ \frac12(\partial_\mu\zeta^{ab}) J^{\ \, \mu}_{C \ \, ab} -(\partial_\mu \xi^a) T^{\ \, \mu}_{C \ \, a}\right] \n\\
&= \int d^4x \left[ \frac12(\partial_\mu \zeta^{ab})S^{\ \, \mu}_{C \ \, ab} - (\partial_\mu \Xi^a - \zeta_{\ \,\, b}^{a}\delta^b_\mu)T^{\ \, \mu}_{C \ \, a} \right], \label{vara}
\end{align}
where we made use of the conservation laws for the canonical currents \eqref{c1} and \eqref{c2}. In order to make  $\delta A$ vanish and thus obtain local Poincar{\'e} invariance, we introduce  $e^\mu_{\ \, a}(x)$ and $\omega_\mu^{\ \, ab}(x)$ as gauge fields in the Lagrangian
 and couple them to the spinors such that 
 \begin{gather}
 \label{w11}
\begin{aligned}
\frac{\partial \mathcal{L}}{\partial e_\mu^{\ \, a}} &\simeq T^{\ \, \mu}_{C \ \, a}, \quad & \delta e_\mu^{\ \, a} &\simeq 
\partial_\mu \Xi^a - \zeta_{\ \,\, b}^{a}\delta^b_\mu , \\
\frac{\partial \mathcal{L}}{\partial \omega_\mu^{\ \, ab}} &\simeq \frac12 S^{\ \, \mu}_{C \ \, ab}, & \delta \omega_\mu^{\ \, ab} &\simeq  \partial_\mu \zeta^{ab} . 
\end{aligned}
\end{gather}
The relations above are supposed to be valid only in the case of weak fields, as a coupling of this form will necessarily modify the canonical currents and their conservation laws which have were used to obtain \eq\eqref{vara}. From this discussion we deduce an important result: if we demand local Poincar{\'e} invariance, then we see that special relativity is not adequate anymore and a deformation of the flat spacetime due to $e^\mu_{\ \, a}(x)$ and $\omega_\mu^{\ \, ab}(x)$ is needed to compensate the change of the action due to the variation of the spinor field. As a consequence, we also have to adjust the derivative operator by replacing it with the covariant derivative. These new fields encode geometrical properties of the spacetime and they indeed represent the gravitational interaction. Such geometry turns out to be the Riemann-Cartan geometry.

We can now relax the assumption of weak gravitational field limit implied in the derivation above. In order to do so, let us assume that in general the Lagrangian density has the following functional dependence
\begin{equation}
 \label{EC_Lagrangian}
  \mathcal{L}=\mathcal{L}[g_{ab}, \gamma_a,\gamma_5, \psi, \partial_\mu\psi, e_\mu^{\ \, a}, \omega_\mu^{\ \, ab}],
\end{equation}
where the Minkowski metric $g_{ab}$ and the Dirac matrices $\gamma_a, \gamma_5$ are constant and defined in the local orthonormal frame. For simplicity in the Lagrangian \eqref{EC_Lagrangian} we omit to write the dependence on the adjoint spinor field. In order for the action 
corresponding to the Lagrangian \eqref{EC_Lagrangian} to be invariant under local Poincar{\'e} transformations, the following condition has to hold~\cite{hehl1976general,hehl1980four,blagojevic2001gravitation}:
\begin{align}
\label{varq}
\frac{\delta\mathcal{L}}{\delta Q}\delta Q+D_\mu \bigg(\frac{\partial\mathcal{L}}{\partial(\partial_\mu Q)}\delta Q + \Xi^\mu\mathcal{L} \bigg)=0,
\end{align}
where $Q=(\psi, e_\mu^{\ a}, \omega_\mu^{\ \, ab})$. It is possible to prove that the general variations of the spinor and gauge fields read
\begin{align}
\delta\psi&= \left(\frac12\zeta^{	ab}f_{ab}-\Xi^a D_a\right)\psi, \\
 \delta e_\mu^{\,\ a}&=D_\mu \Xi^a - \zeta_{\ \,\, b}^{a}e_\mu^{\,\ b}{+\Xi^b F_{b \mu}^{\ \  a}},\\
 \delta \omega_\mu^{\ \,\, ab}&= D_\mu\zeta^{ab}{+\Xi^c R_{c \mu}^{\,\ \ ab}},
\end{align}
(cf. with  \eqs\eqref{functvar} and \eqref{w11}). From \eq\eqref{varq}, requiring that the functions multiplying the independent quantities $D_\mu\Xi^a$,  $D_\mu\zeta^{ab}$,  $\Xi^a, \ \zeta^{ab}$ vanish, after using the equations of motion for $\psi$ we obtain
\begin{align}
 e\,  T^{\ \, \mu}_{C \ \, a}&=\frac{\partial \mathcal{L}}{\partial e_\mu^{\ \, a}}=\frac{\partial\mathcal{L}}{\partial(\partial_\mu\psi)}D_a\psi -e^\mu_{\ \, a}\mathcal{L},\label{ememem}\\
 e\, S^{\ \,\, \mu}_{C \ \, ab}&=2\frac{\partial \mathcal{L}}{\partial \omega_\mu^{\ \, ab}}=\frac{\partial\mathcal{L}}{\partial(\partial_\mu\psi)}f_{ab}\psi,\label{cons22}\\
D_\mu(e\, T^{\ \, \mu}_{C \ \, a})&=F_{a \mu}^{\ \ \, b}e\ T^{\ \, \mu}_{C \ \, b}+\frac12 R_{a \mu}^{\ \ \, \, bc}e\ S^{\ \,\, \mu}_{C \ \,\, bc},\label{T_kin_EC}\\
D_\mu (e\, S^{\ \,\, \mu}_{C \ \, ab})&= e T_{C\, [ba]},
\label{S_kin_EC}
\end{align}
respectively, where $e=\text{det}(e_\mu^{\ \, a})$.   
We can ensure the validity of Eqs.~\eqref{ememem}-\eqref{S_kin_EC} by applying the so-called minimal coupling when generalizing the special-relativistic Lagrangian to the Einstein-Cartan theory
\begin{equation}
 \mathcal{L}(\psi,\delta_a^\mu\partial_\mu\psi)
 \rightarrow e\, 
 \mathcal{L}(\psi, e_{\ \, a}^\mu D_\mu\psi).
\end{equation}
We stress that in the Einstein-Cartan theory the currents which arise by taking the variations with respect to the gauge fields $e_\mu^{\ \, a}$ and $\omega_\mu^{\ \, ab}$ reduce in flat spacetime to the canonical currents.  

We conclude this section by mentioning that it is in principle possible to generalize also quadratic Lagrangians (such as the squared Dirac Lagrangian) to curved spacetime and thus connect the HW tensors to the Einstein-Cartan theory \cite{Seitz:1985ce,Obukhov:2014fta}.

\section{Conclusions}
\label{concsec}

The relativistic decomposition of the total angular momentum is an old problem which embraces many branches of physics. In this review we focused on some formal aspects and applications which are decades old and on some others which have recently attracted considerable attention.
{We showed in different contexts which choice appears to be a physical one. In particular, we reviewed some of the latest results  regarding the description of spin dynamics in relativistic fluids in relation to the physics of the QGP in heavy-ion collisions. In this case, the HW pseudo-gauge has important advantages, namely the spin tensor is not conserved only when nonlocal particle scatterings are considered, and orbital angular momentum can be converted into spin through the antisymmetric part of the energy-momentum tensor. Eventually, when global equilibrium is reached, the HW spin tensor is conserved and no orbital-to-spin angular momentum conversion can occur. 
Moreover, we discussed how different definitions of the relativistic center of inertia are related to the pseudo-gauge transformations. In this context,  we showed that, unlike in the massless case, for massive particles it is always possible to define the spin in a covariant way using the HW, GLW or KG global spin (at least for free fields or in the case of local interactions). The fact that the definition of the spin of a massless particle is inherently noncovariant leads to the side-jump effect. Finally, we discussed the Einstein-Cartan theory, an extension of general relativity which allows a physical definition through geometry of an asymmetric energy-momentum tensor and a spin tensor which reduce to the conventional canonical currents in flat spacetime. 

The ultimate question one would like to address is how to extract information from experiments on which pseudo-gauge describes the system and whether it is unique. A possible answer can be found exploiting  the fact that, in general, the expectation value of an observable onto a state for which local equilibrium can be defined, does depend on the pseudo-gauge. Furthermore, in relativistic spin hydrodynamics the values of the fields may be different with respect to which pseudo-gauge one uses to decompose the total angular momentum. In particular, in the Belinfante case, one does not need to introduce the spin potential as an additional dynamical field and the spin dynamics may be different than in other pseudo-gauges.  On the other hand, in the context of gravitational physics, the way one is supposed to measure energy, momentum and spin densities is through spacetime geometry. 
}

One may expect that the formalism and the problems covered in the present review will be relevant in the near future since they are shared in different fields, some of which are and will be under active experimental investigation. In heavy-ion collisions the development of dissipative spin hydrodynamics and quantum kinetic theory appears to be an urgent task in order to understand the nontrivial dynamics of polarization, especially in light of the recent and future experimental program \cite{Becattini:2020ngo}. Some important questions one would like to address are  whether spin can equilibrate fast enough for time scales relevant to nuclear collisions and how this nonequilibrium dynamics can modify the expression for spin polarization commonly used to describe the Lambda global polarization data. 

Understanding how angular momentum can be split into an orbital and spin part is also crucial for the description of gauge fields. In addition to the complications discussed in this paper, there is the question of whether it is possible to find a gauge-invariant way to decompose the total angular momentum. A fundamental description of the spin and orbital angular momentum of light is still controversial, see, e.g., works related to optics in
 Refs.~\cite{cameron2012optical,bliokh2014conservation}. Recently, a gauge-invariant measure of the spin of the photon called zilch current has been studied also in the context of quantum kinetic theory and nuclear collisions~\cite{Chernodub:2018era,Huang:2020kik}.
Furthermore, in hadron physics, the angular momentum decomposition is of utmost importance to understand the contribution of quarks and gluons to the spin of the nucleon~\cite{Leader:2013jra}. For related works about the angular momentum decomposition with a focus on chiral physics, see Refs. \cite{Fukushima:2018osn,Fukushima:2020qta} in which connections to nuclear collisions are also discussed. Addressing the problem of the nucleon 
 spin will also be at the core of the experimental effort of the future electron-ion collider \cite{Boer:2011fh,Accardi:2012qut}. 
 Finally, we would like to mention applications of spin dynamics in cosmology, where the Einstein-Cartan theory is often taken as a starting point \cite{Obukhov:1987yu,Poplawski:2010kb}, and in  condensed matter systems like, e.g., spintronics \cite{takahashi2016spin}.

\section*{Acknowledgments}

We thank F. Becattini, W. Florkowski, D. H. Rischke, R. Ryblewski, A. Sadofyev, X.-L. Sheng, L. Tinti, G. Torrieri, D. Wagner and Q. Wang for enlightening discussions and fruitful collaborations. We are particularly grateful to A. Sadofyev also for extremely valuable comments on the manuscript. 
This work was supported by the
Deutsche Forschungsgemeinschaft (DFG, German Research Foundation)
through the Collaborative Research Center CRC-TR 211 ``Strong-inter{\-}action matter
under extreme conditions'' -- project number 315477589 - TRR 211.
E.S. acknowledges support also by the COST Action CA15213 ``Theory of hot matter and relativistic heavy-ion collisions (THOR)".

\begin{appendix}

\section{Lorentz transformation properties of hyper{\-}surface-in{\-}tegrated quantities}
\label{applor}

In this appendix we show that, given a generic rank-$(n+1)$ tensor $\hat{B}^{\lambda \mu_1\cdots \mu_n} (x)$, the quantity defined as
\begin{equation}
\label{intap}
 \hat{\mathcal{B}}^{\mu_1\cdots \mu_n}=\int_\Sigma d \Sigma_\lambda \hat{B}^{\lambda \mu_1\cdots \mu_n} (x)
\end{equation}
transforms as a rank-$n$ tensor only if $\hat{B}^{\lambda \mu_1\cdots \mu_n}(x)$ is a conserved quantity, i.e., $\partial_\lambda \hat{B}^{\lambda \mu_1\cdots \mu_n}(x)=0$, and suitable boundary conditions are fulfilled. This also means that the quantity in Eq. \eqref{intap} is independent of the choice of the hypersurface integration. The correct properties under Lorentz transformations are not in general guaranteed because, in the integrated quantity
 $\hat{\mathcal{B}}^{\mu_1\cdots \mu_n}$, only the fields in the integrand have to be transformed and the integration hypersurface must be unchanged. In other words, the experimenter is not transformed. {This is the point of view of an active transformation which we will adopt in the proof below. 
We could instead perform a passive transformation, namely change the observer, i.e., the hypersurface, and leave the fields unchanged. The proof is independent of whether we do an active or passive transformation.}
{Since Eq. \eqref{intap} is expressed in a covariant way in terms of a scalar product, if we changed both the fields and the hypersurface together, it would look like it has the proper transformation behavior. However, this is not what one would demand from a Lorentz transformation, as either the fields or the observer should be changed.}
   To show this, we closely follow the proof given in Ref. \cite{Coleman} and we stress that it is valid for both quantum and classical fields.

Let us choose a hypersurface at $x^0=0$, hence  $d\Sigma_\lambda=d^4 x \, \delta (\nnn \cdot x) \nnn_\lambda$, where $\nnn_\lambda=(1,{\bf 0})$. Thus, Eq. \eqref{intap} becomes
\begin{align}
\label{bbapp}
\hat{\mathcal{B}}^{\mu_1\cdots \mu_n} =& \int d^3 x\,  \hat{B}^{0 \mu_1\cdots \mu_n} (x) \n \\
=&\int d^4 x \, \delta (\nnn \cdot x) \nnn_\lambda \hat{B}^{\lambda \mu_1\cdots \mu_n} (x) \n\\
=& \int d^4 x \, [\partial_\lambda\theta(\nnn \cdot x) ] \hat{B}^{\lambda \mu_1\cdots \mu_n} (x),
\end{align}
where we introduced the step function such that $\theta (\nnn\cdot x)=1$ for $\nnn\cdot x \geq 0$ and $\theta (\nnn\cdot x)=0$ for $\nnn\cdot x < 0$ and used the relation
\begin{equation}
\label{eqtrr}
\nnn_\lambda \delta (\nnn \cdot x) = \partial_\lambda \theta (\nnn\cdot x).
\end{equation}
Note that in Eq. \eqref{bbapp} we clearly cannot neglect boundary terms in the temporal direction after integrating by parts, as $\theta (\nnn\cdot x)=1$ when $x^0\to +\infty$.
We now study how $\hat{\mathcal{B}}^{\mu_1\cdots \mu_n}$ transforms under a Lorentz transformation $\Lambda^\mu_\nu$. As previously discussed, we perform a transformation of the fields under the integral. Since $\hat{B}^{\lambda \mu_1\cdots \mu_n} (x)$ is a tensor, it will transform as
\begin{equation}
 \hat{B}^{\,\prime\,\lambda \mu_1\cdots \mu_n} (x) =  \Lambda^\lambda_\rho \Lambda^{\mu_1}_{\sigma_1} \cdots \Lambda^{\mu_n}_{\sigma_n} \hat{B}^{\rho \sigma_1\cdots \sigma_n} (\Lambda^{-1}x).
\end{equation}
Therefore, the act of the Lorentz transformation on Eq. \eqref{bbapp} is given by
\begin{equation}
\label{appbef}
\hat{\mathcal{B}}^{\,\prime\,\mu_1\cdots \mu_n} = \int d^4 x \, \delta (\nnn \cdot x) \nnn_\lambda \Lambda^\lambda_\rho \Lambda^{\mu_1}_{\sigma_1} \cdots \Lambda^{\mu_n}_{\sigma_n} \hat{B}^{\rho \sigma_1\cdots \sigma_n} (\Lambda^{-1}x).
\end{equation}
We now change the integration variable $x^\prime=\Lambda^{-1} x$ and define $\nnn^\prime = \Lambda^{-1}\nnn$ so that $\nnn\cdot x= \nnn^\prime \cdot x^\prime$. Hence, Eq. \eqref{appbef} becomes
\begin{align}
\label{tr111}
\hat{\mathcal{B}}^{\,\prime\,\mu_1\cdots \mu_n} ={}& \int d^4 x^\prime \, \Lambda_{\lambda\alpha}\nnn^{\prime\,\alpha} \Lambda^\lambda_\rho \Lambda^{\mu_1}_{\sigma_1} \cdots \Lambda^{\mu_n}_{\sigma_n} \hat{B}^{\rho \sigma_1\cdots \sigma_n}   (x^\prime) \n \\
&\times \delta (\nnn^\prime \cdot x^\prime) \n \\
={}& \Lambda^{\mu_1}_{\sigma_1} \cdots \Lambda^{\mu_n}_{\sigma_n} \int d^4 x^\prime \, \delta (\nnn^\prime \cdot x^\prime)\nnn^{\prime}_\rho   \hat{B}^{\rho \sigma_1\cdots \sigma_n}   (x^\prime)  \n \\
={}& \Lambda^{\mu_1}_{\sigma_1} \cdots \Lambda^{\mu_n}_{\sigma_n} \int d^4 x \, \delta (\nnn^\prime \cdot x)\nnn^{\prime}_\rho  \hat{B}^{\rho \sigma_1\cdots \sigma_n}   (x) ,
\end{align}
where in the first equality we made use of the invariance of $d^4 x$ under the transformation $\Lambda$, in the second one $\Lambda_{\lambda\alpha}\Lambda^{\lambda}_\rho = g_{\alpha\rho}$, and in the last one we only relabeled the integration variable by dropping the prime index. On the other hand, if $\hat{\mathcal{B}}^{\,\prime\,\mu_1\cdots \mu_n}$ transforms like a rank-$n$ tensor, we must also have 
\begin{align}
\label{tr222}
\Lambda^{\mu_1}_{\sigma_1} \cdots \Lambda^{\mu_n}_{\sigma_n}\hat{\mathcal{B}}^{ \sigma_1\cdots \sigma_n}={}& \Lambda^{\mu_1}_{\sigma_1} \cdots \Lambda^{\mu_n}_{\sigma_n} \int d^4 x \, \delta (\nnn \cdot x) \n \\
{}&\times\nnn_\rho  \hat{B}^{\rho \sigma_1\cdots \sigma_n}   (x) .
\end{align}
Therefore, to conclude the proof, we only need to find the conditions for which Eqs. \eqref{tr111} and \eqref{tr222} are equal, which is the same as requiring that the following difference vanishes:
\begin{align}
&\hat{\mathcal{B}}^{\mu_1\cdots \mu_n} - (\Lambda^{-1})^{\mu_1}_{\sigma_1} \cdots (\Lambda^{-1})^{\mu_n}_{\sigma_n}\hat{\mathcal{B}}^{ \,\prime\,\sigma_1\cdots \sigma_n } \n \\ 
&= \int d^4 x\, \partial_\rho [\theta (\nnn\cdot x) - \theta (\nnn^\prime\cdot x)] \hat{B}^{\rho \mu_1\cdots \mu_n}   (x) \n\\
&= -\int d^4 x\, [\theta (\nnn\cdot x) - \theta (\nnn^\prime\cdot x)]\partial_\rho  \hat{B}^{\rho \mu_1\cdots \mu_n}   (x),
\label{diftrvec}
\end{align}
where we used the relation \eqref{eqtrr} and, in the last step, we integrated by parts and safely neglected boundary terms in the spatial as well as in the temporal direction, since in the far future or far past we have $\theta (\nnn\cdot x)=\theta (\nnn^\prime\cdot x)$. Finally, Eq. \eqref{diftrvec} vanishes only if 
\begin{equation}
\label{conserva}
\partial_\lambda \hat{B}^{\lambda \mu_1\cdots \mu_n}(x)=0,
\end{equation}
which is what we wanted to proof. Note that what we also showed is that the integral in Eq. \eqref{bbapp} is independent of the choice of the hypersurface using the divergence theorem, provided that Eq. \eqref{conserva} holds and suitable boundary conditions are fulfilled. In fact, the second line of Eq. \eqref{diftrvec} is the difference of the integral of $\hat{B}^{\rho \mu_1\cdots \mu_n}   (x)$ calculated at two different hypersurfaces with normal vectors $\nnn_\mu$ and $\nnn_\mu^\prime$, respectively, which is transformed to a volume integral in the last line. What discussed in this section is clearly not restricted to the form of the hypersurface chosen for the proof and can be easily generalized to a generic shape. 
Consider a region of spacetime enclosed between two space-like hypersurfaces $\Sigma_1$, $\Sigma_2$ corresponding to two different values of the parameter $\mathfrak{t}$ used for the foliation of the spacetime, $\mathfrak{t}_1$, $\mathfrak{t}_2$, respectively ($\mathfrak{t}$ can be, e.g., $x^0$). Using the divergence theorem and the fact that we can neglect terms at the boundaries, we have
\begin{equation}
\int_{\Sigma_1} d \Sigma_\lambda \hat{B}^{\lambda \mu_1\cdots \mu_n}-\int_{\Sigma_2} d \Sigma_\lambda \hat{B}^{\lambda \mu_1\cdots \mu_n} = \int_V d V  \partial_\lambda \hat{B}^{\lambda \mu_1\cdots \mu_n},
\end{equation}
where $V$ is the four-dimensional volume. The right-hand side of the equation above vanishes if \eqref{conserva} is valid.

We conclude that, since the energy-momentum and  total angular momentum tensors are conserved quantities, the associated charges in Eqs. \eqref{totp} and \eqref{totj} transform properly as tensors under Lorentz transformations. On the other hand, since the canonical spin tensor is not conserved [see Eq. \eqref{nccan}], the associated global spin in Eq. \eqref{totcanspin} will not transform as a tensor.

\section{Matrix element of the Pauli-Lubanski vector}
\label{apppp2}

In the first part of this appendix we show the details of the derivation of \eq\eqref{whatweprove0}. We first introduce the plane-wave expansion of the Dirac field 
\begin{align}
\label{plane}
\ps(x)=&\frac{1}{\sqrt{2(2\pi\hbar)^3}}\sum_{r=1}^2 \int \frac{d^3k}{k^0}\, \left[u_r({ k}) e^{-\frac{i}{\hbar}k\cdot x} a_r({ k })  \right.\n\\
&\left.+ v_r({k}) e^{\frac{i}{\hbar}k\cdot x} b^\dagger_r({k})\right],
\end{align}
where $u_r({k})$ and $v_r({k})$ are the standard spinors for particles and antiparticles, respectively. Moreover, $a_r({ k })$ is the annihilation operator and $b^\dagger_r({k})$ is the creation operator of respectively a particle and antiparticle state with momentum ${ k^\mu }$ and spin projection $r$. We start by proving that
\begin{align}
&\bra{ \ppm^\prime,s^\prime} \hat{S}^\mu \ket{\ppm,s} \n\\ 
={}&-\frac1{2m}\epsilon^{\mu\nu\alpha\beta}\bra{ \ppm^\prime,s^\prime}  \hat{P}_\nu\J_{\alpha\beta}\ket{\ppm,s}\n\\
={}&-\frac1{2m}\epsilon^{\mu\nu\alpha\beta}\ppm_\nu\int d^3x\, \bra{ \ppm^\prime,s^\prime} \ps^\dagger(x) \frac\hbar2 \sigma_{\alpha\beta} \ps(x)\ket{\ppm,s},
\label{whatweprove00}
\end{align}
where 
\begin{equation}
|\ppm,s\rangle= a_s^\dagger(\ppm) \ket{0}.
\end{equation}
is the one-particle state of momentum $\ppm$ and spin projection $s$
(for antiparticle states the proof would follow similar steps).
It is convenient to note that
\begin{equation}
\label{asdasd2}
\epsilon^{\mu\nu\alpha\beta} \hat{P}_\nu\J_{\alpha\beta} = \epsilon^{\mu\nu\alpha\beta} \J_{\alpha\beta} \hat{P}_\nu,
\end{equation}
since the total charges obey the Poincar\'{e} algebra
\begin{equation}
[\hat{P}_\nu,\J_{\alpha\beta}]=i(g_{\beta\nu}\hat{P}_\alpha - g_{\alpha\nu}\hat{P}_\beta).
\end{equation}
Using \eqs\eqref{totj}, \eqref{asdasd2} and $\hat{P}^\mu \ket{ \ppm,s} = \ppm^\mu \ket{ \ppm,s}$, we get
\begin{align}
&\bra{ \ppm,s }\hat{S}^\mu \ket{ \ppm,s}  \n\\
={}&-\frac{1}{2m} \epsilon^{\mu\nu\alpha\beta} \mathfrak{p}_\nu\int d^3x \bra{ \ppm^\prime ,s^\prime }\psi^\dagger \left[ 2i\hbar x_\alpha \partial_{\beta} + \frac \hbar 2 \sigma_{\alpha\beta}\right] \psi \ket{ \ppm,s}.
\label{francesco2}
\end{align}
We now calculate the orbital part
\begin{align}
&-\frac{i\hbar}{m}\epsilon^{\mu\nu\alpha\beta} \mathfrak{p}_\nu  \int d^3x \bra{ \ppm^\prime ,s^\prime } \psi^\dagger (x)x_\alpha \partial_\beta \psi(x) \ket{ \ppm,s} \n\\
={}&-\frac{1}{{2m(2\pi\hbar)^3}} \epsilon^{\mu\nu\alpha\beta}\ppm_\nu \int d^3x \int \frac{d^3 k d^3 k^\prime}{k_0k^\prime_0} x_\alpha k_\beta  \n\\
{}&\times u^\dagger_{r^\prime}(k^\prime) u_r(k) e^{\frac i\hbar (k^\prime-k)\cdot x} \bra{ \ppm^\prime s^\prime }a^\dagger_{r^\prime}(k^\prime)a_{r}(k) \ket{ \ppm , s }  \n\\
={}&-\frac{1}{{2m(2\pi\hbar)^3}} \epsilon^{\mu\nu\alpha\beta}\ppm_\nu \int d^3x \int \frac{d^3 k d^3 k^\prime}{k_0k^\prime_0} x_\alpha k_\beta u^\dagger_{r^\prime}(k^\prime) u_r(k)  \n \\ 
{}&\times e^{\frac i\hbar (k^\prime-k)\cdot x} \ppm_0 \ppm^\prime_0 \delta^{(3)}(\mathbf{k}-\pp)\delta^{(3)}(\mathbf{k}^\prime-\pp^\prime) \delta_{rs}\delta_{r^\prime s^\prime}  \n\\
%={}&-\frac{1}{{2m(2\pi\hbar)^3}} \epsilon^{\mu\nu\alpha\beta}\ppm_\nu \ppm_\beta \int d^3x  x_\alpha   u^\dagger_{s^\prime}(\ppm^\prime) u_s(\ppm) e^{\frac i\hbar (\ppm^\prime -\ppm)\cdot x} \n \\
={}&0.
\label{appb11}
\end{align}
In deriving the result above  we made use of \eq\eqref{plane}, the antisymmetry of the Levi-Civita tensor together with the relation
\begin{align}
\label{adaggera}
&\bra{\ppm^\prime s^\prime }a^\dagger_{r^\prime}(k^\prime)a_{r}(k) \ket{ \ppm , s } \n\\
={}&\bra{ 0 }a_{s^\prime} (\ppm^\prime) a^\dagger_{r^\prime}({k}^\prime)a_{r}({k}) a^\dagger_s (\ppm) \ket{0 }\n \\
={}&\ppm_0 \ppm^\prime_0 \delta^{(3)}(\mathbf{k}-\pp)\delta^{(3)}(\mathbf{k}^\prime-\pp^\prime)\delta_{rs}\delta_{r^\prime s^\prime},
\end{align}
which follows from 
\begin{equation}
\{ a_r({k}), a_s^\dagger(\ppm)\} = \ppm^0 \delta(\mathbf{k}-\pp)\delta_{rs}
\end{equation}
and $a_r({k}) \ket{ 0}=0$. Using \eq\eqref{appb11}, \eq\eqref{francesco2} simplifies to \eq\eqref{whatweprove00}. 

Consider now the Pauli-Lubanski vector written in terms of the canonical and HW pseudo-gauge. Using \eqs\eqref{tcan} and \eqref{thw} and following similar steps which lead to \eq\eqref{appb11}, one can prove that the orbital parts cancel when taking the matrix element of one-particle states. For the canonical global spin contribution, from \eq\eqref{scan} we have
\begin{align}
\Sp_C^{i0}={}& 0,\n\\
\Sp_C^{ij}={}& \int d^3x\, \ps^\dagger(x) \frac\hbar2 \sigma^{ij} \ps(x),
\end{align}
where we performed the spacetime integration at constant $x^0$.
Using $\bar{u}_{r^\prime}(k)\gamma^\mu u_r(k)= 2 k^\mu \delta_{rr^\prime}$, one can prove that
 the contribution to \eq\eqref{whatweprove00} with $\alpha=l$, $\beta=0$ vanishes ($l=1,2,3$), which in turn implies 
\begin{equation}
\bra{ \ppm^\prime,s^\prime} \hat{S}^\mu\ket{\ppm,s}  
=-\frac1{2m}\epsilon^{\mu\nu\alpha\beta}\bra{ \ppm^\prime,s^\prime}\hat{P}_\nu\Sp_{C,\alpha\beta}\ket{\ppm,s} .
\label{whatweprove0123}
\end{equation}
Furthermore, following similar steps as in \eq\eqref{appb11}, one can show that the contribution to the Pauli-Lubanski vector given by the expectation value of the second term in the last line of \eq\eqref{stothw} cancels when taking the matrix element. Thus, we also have 
\begin{align}
\bra{ \ppm^\prime,s^\prime} \hat{S}^\mu\ket{\ppm,s} -\frac1{2m}\epsilon^{\mu\nu\alpha\beta}\bra{ \ppm^\prime,s^\prime}\hat{P}_\nu\Sp_{HW,\alpha\beta}\ket{\ppm,s} .
\label{whatweprove01234}
\end{align}

We conclude the appendix by proving \eq\eqref{seqpi}. Using \eq\eqref{eqwig22} and plugging \eq\eqref{plane} into \eq\eqref{wtrans}, we obtain \cite{Becattini:2020sww,Tinti:2020gyh}
\begin{align}
&\int d\Sigma_\lambda \,p^\lambda	\hat{W}_{\kappa\chi} (x,p)  \n\\
={}& \int d^3x  \,p^0	\hat{W}_{\kappa\chi} (x,p) \n\\
={}& \delta (p^2-m^2)\sum_{r,r^\prime}\Big[\theta(p^0)a^\dagger_{r}(p) a_{r^\prime}(p)u_{\kappa, r^\prime}(p)\bar{u}_{\chi, r}(p) \n\\
&+\theta (-p^0)b^\dagger_{r^\prime}(-p)b_{r}(-p)v_{\kappa,r^\prime}(-p)\bar{v}_{\chi,r}(-p)\Big]
\end{align}
which shows that the momentum $p^\mu$ is on-shell. After taking the matrix element on one-particle states and using \eq\eqref{adaggera},
we get \eq\eqref{seqpi}.

\end{appendix}

%
% BibTeX users please use
% \bibliographystyle{}
% \bibliography{}
%
% Non-BibTeX users please use
\bibliography{biblio_paper_review}{}
\bibliographystyle{epj.bst}
%
%\begin{thebibliography}{}
%%
%% and use \bibitem to create references.
%%
%\bibitem{RefJ}
%% Format for Journal Reference
%Author, Journal \textbf{Volume}, (year) page numbers.
%% Format for books
%\bibitem{RefB}
%Author, \textit{Book title} (Publisher, place year) page numbers
%% etc
%\end{thebibliography}

\end{document}